\documentclass[a4paper,fleqn]{cas-sc}
\usepackage{makecell}
\usepackage{diagbox}
\usepackage{lscape}
\usepackage[authoryear]{natbib}
\usepackage{wrapfig}

\usepackage{booktabs}
\usepackage{longtable}
\usepackage{array}
\usepackage{enumitem}
\usepackage{adjustbox}
\usepackage{pifont} 

\usepackage[ruled,vlined]{algorithm2e}
\usepackage{algpseudocode} 
\usepackage{graphicx}
\graphicspath{ {Figures/} }
\usepackage{caption}
\usepackage{subcaption}

\def\tsc#1{\csdef{#1}{\textsc{\lowercase{#1}}\xspace}}
\tsc{WGM}
\tsc{QE}
\tsc{EP}
\tsc{PMS}
\tsc{BEC}
\tsc{DE}

\begin{document}
\let\WriteBookmarks\relax
\def\floatpagepagefraction{1}
\def\textpagefraction{.001}
\shorttitle{MetaInfoSci}

\title [mode = title]{MetaInfoSci: An Integrated Web Tool for Scholarly Data Analysis}                      
\author[1,2]{Kiran Sharma}\corref{cor1}
\ead{kiran.sharma@bmu.edu.in}

\author[3]{Parul Khurana}
\author[1,2]{Ziya Uddin}
%
\address{School of Engineering \& Technology, BML Munjal University, Gurugram, Haryana-122413, India }
\address{Center for Advanced Data and Computational Science, BML Munjal University, Gurugram, Haryana-122413, India }
\address{School of Computer Applications, Lovely Professional University, Phagwara, Punjab-144411, India }
\cortext[cor1]{Corresponding author}


\begin{abstract}
The exponential increase in academic publications has made it increasingly difficult for researchers to remain up to date and systematically synthesize knowledge scattered across vast and fragmented research domains. Literature reviews, particularly those supported by bibliometric methods, have become essential in organizing prior findings and guiding future research directions. While numerous tools exist for bibliometric analysis and network science, there is currently no single platform that integrates the full range of features from both domains. Researchers are often required to navigate multiple software environments, many of which lack customizable visualizations, cross-database integration, and AI-assisted result summarization. Addressing these limitations, this study introduces MetaInfoSci (\url{www.metainfosci.com}), a comprehensive, web-based platform designed to unify bibliometric, scientometric, and network analytical capabilities. The platform supports tailored query design, merges data from diverse sources, enables rich and adaptable visual outputs, and provides automated, AI-driven summaries of analytical results. This integrated approach aims to enhance the accessibility, efficiency, and depth of scientific literature analysis for scholars across disciplines.

\end{abstract}

%
%
%

\begin{keywords}
Bibliometrics  \sep Scientometrics \sep Network analysis \sep Custom visualization \sep Custom query handling \sep AI-enabled summary

\end{keywords}

\maketitle


\section{Introduction}
The exponential growth of academic publications has made it increasingly unfeasible for scholars and practitioners to stay current with all relevant developments in their fields. The dominant focus on empirical studies has resulted in extensive yet fragmented research streams, making it difficult to systematically consolidate and interpret existing knowledge. As a result, literature reviews have become essential for structuring previous findings and shaping the direction of future research \citep{briner2012systematic}. Consequently, literature reviews play a pivotal role in organizing prior findings, guiding future research, and offering evidence-based insights into both scholarly and professional practices \citep{rousseau2012oxford}.

To manage the deluge of research outputs, scholars are increasingly relying on systematic review techniques, particularly bibliometrics, which offers a structured, objective, and reproducible methodology for assessing scientific literature \citep{sharma2021growth}. Bibliometric analysis enables researchers to trace the development of knowledge domains over time, identify influential scholars and institutions, and detect thematic trends and shifts in disciplinary boundaries \citep{khurana2024growth}. This quantitative approach has gained traction across disciplines, yet it remains complex and often inaccessible due to its reliance on multiple analytical tools and specialized skills \citep{guler2016scientific}.

As bibliometric data becomes increasingly complex and interconnected, network science has emerged as a powerful complementary framework. Network science, rooted in graph theory and statistical physics, offers a set of tools to model and analyze relationships among entities—such as co-authorship, citation networks, and keyword co-occurrence in bibliometric datasets. By examining structural properties such as centrality, modularity, and clustering, network-based bibliometrics uncovers hidden patterns of influence and collaboration that traditional statistics may overlook \citep{posfai2016network}.

Recent literature underscores the importance of network approaches in bibliometrics. For example, Waltman et al. (2020) advocate for science mapping techniques that visualize the intellectual structure of disciplines through citation and co-citation networks. Similarly, \cite{chen2016citespace} emphasizes the role of temporal and dynamic network analysis in tracking the evolution of research topics over time. These advances not only enrich bibliometric reviews but also enable more nuanced interpretations of disciplinary growth and interconnectivity.

Moreover, the integration of automated workflows and network analysis is accelerating the ability to conduct large-scale, reproducible bibliometric studies. Tools such as VOSviewer, CiteSpace, and Biblioshiny (a Shiny interface for Bibliometrix) offer interactive visualization and modular analysis, expanding the scope and accessibility of bibliometric research \citep{van2017citation, chen2004searching,aria2017bibliometrix}. The application of community detection algorithms and network centrality measures allows for identifying emerging research fronts and influential knowledge hubs within scientific ecosystems.

\begin{figure}[!h]
    \centering
\includegraphics[width=0.9\linewidth]{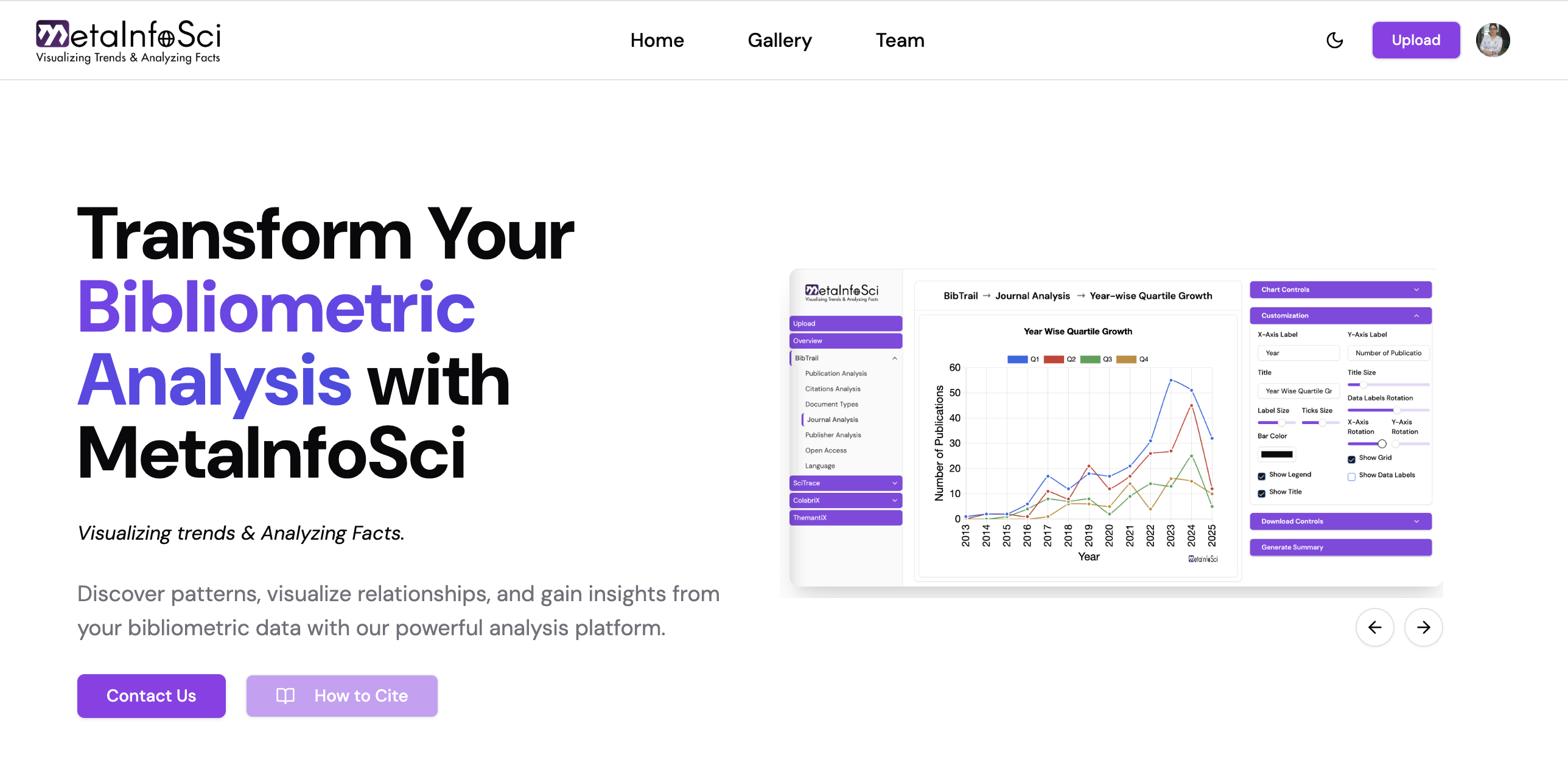} 
\caption{MetaInfoSci home page.}
\label{fig:home}
\end{figure}

Currently,there are several bibliometric tools and network analysis platforms available for analyzing scientific publications. However, there is currently no single platform that offers comprehensive features from both domains in one place. Scholars need to become familiar with different platforms to analyze their publications. Additionally, existing bibliometric software lacks customizable visualization options, merging different databases \citep{khurana2022comparative} and AI summary of results, a limitation that MetaInfoSci (\url{www.metainfosci.com}) aims to overcome (see Fig~\ref{fig:home}).

\subsection{Research objectives}
\begin{itemize}
    \item Develop a specialized web-based tool that integrates bibliometric, scientometric, and network analysis for comprehensive scholarly data analysis.

\item Provide unified analytics with features like customizable visualizations, tailored query design, merged databases, and AI-powered result summarization.
\end{itemize}
~~~~~~~~~~~~~~~~~~~~~~~~~~~~~~~~~~~~~~~~~~~~~~~~~~~~~~~~~~~~~~~~~~~~~~~

~~~~~~~~~~~~~

\section{Literature Review}
Table~\ref{Tab:Literature} demonstrates the existing tools for scholarly data analysis along with their key features and limitations.

\begin{table}[]
\caption{List of existing Bibliometrics, citations and network analysis tools to analyze scholarly publishing data.}
\label{Tab:Literature}

\begin{tabular}{|lllll|}
\hline
\multicolumn{5}{|l|}{\textbf{Bibliometric analysis only}}                                   \\ \hline
\multicolumn{1}{|l|}{\textbf{Sr.}} & \multicolumn{1}{l|}{\textbf{Name}}                             & \multicolumn{1}{l|}{\textbf{Type}}                                            & \multicolumn{1}{l|}{\textbf{Key Features}}                                                                                                                & \textbf{Limitations}                                                                                                                                     \\ \hline
\multicolumn{1}{|l|}{1}            & \multicolumn{1}{l|}{\begin{tabular}[c]{@{}l@{}}Biblioshiny\\  \citep{aria2017bibliometrix}\end{tabular}
}                               & \multicolumn{1}{l|}{\begin{tabular}[c]{@{}l@{}}Web Tool\\ GUI\end{tabular}}   & \multicolumn{1}{l|}{\begin{tabular}[c]{@{}l@{}} \textbullet Shiny-based GUI\\ \textbullet Bibliometrix integration\\ \textbullet User-friendly statistical\\ visualization\end{tabular}} & \begin{tabular}[c]{@{}l@{}}\textbullet Limited customization\\ \textbullet Requires R setup\\ \textbullet Lacks real-time bibliometric updates\end{tabular}                            \\ \hline
\multicolumn{1}{|l|}{2}            & \multicolumn{1}{l|}{\begin{tabular}[c]{@{}l@{}}BibExcel\\ \citep{persson2009use}\end{tabular}
}                                  & \multicolumn{1}{l|}{Software}                                                 & \multicolumn{1}{l|}{\begin{tabular}[c]{@{}l@{}}\textbullet Bibliographic data\\ conversion\\ \textbullet Excel output\\ \textbullet Network preparation\\ \textbullet Citation analysis\end{tabular}}     & \begin{tabular}[c]{@{}l@{}}\textbullet Not user-friendly for large datasets\\ \textbullet Lacks real-time updates\\ \textbullet Limited interactivity\end{tabular}                           \\ \hline
\multicolumn{1}{|l|}{3}            & \multicolumn{1}{l|}{\begin{tabular}[c]{@{}l@{}}ScientoPy\\ \citep{ruiz2019software}\end{tabular}
}                                 & \multicolumn{1}{l|}{Software}                                                 & \multicolumn{1}{l|}{\begin{tabular}[c]{@{}l@{}}\textbullet Trend analysis\\ \textbullet Performance metrics\\ \textbullet Bibliographic coupling\\ \textbullet Statistical analysis\end{tabular}}         & \begin{tabular}[c]{@{}l@{}}\textbullet Limited visualization options\\ \textbullet Requires manual data preparation\\ \textbullet Lacks cloud-based accessibility\end{tabular}               \\ \hline
\multicolumn{1}{|l|}{4}            & \multicolumn{1}{l|}{\begin{tabular}[c]{@{}l@{}}pyBibX\\ \citep{pereira2025pybibx}\end{tabular}
}                                    & \multicolumn{1}{l|}{\begin{tabular}[c]{@{}l@{}}Python\\ Library\end{tabular}} & \multicolumn{1}{l|}{\begin{tabular}[c]{@{}l@{}}\textbullet Python library for\\ bibliometric analysis\\ \textbullet Machine learning\\ integration\end{tabular}}                  & \begin{tabular}[c]{@{}l@{}}\textbullet Requires Python knowledge\\ \textbullet Steep learning curve\\ \textbullet Lacks user-friendly UI\end{tabular}                                        \\ \hline
\multicolumn{5}{|l|}{\textbf{Citations Analysis + Bibliometric Networks}}\\ \hline
\multicolumn{1}{|l|}{5}            & \multicolumn{1}{l|}{\begin{tabular}[c]{@{}l@{}}VOSviewer\\ \citep{van2009software}\end{tabular}
}         & \multicolumn{1}{l|}{Software}                                                 & \multicolumn{1}{l|}{\begin{tabular}[c]{@{}l@{}}\textbullet Co-occurrence analysis\\ \textbullet Bibliographic coupling\\ \textbullet Network visualization\\ \textbullet Clustering\end{tabular}}         & \begin{tabular}[c]{@{}l@{}}\textbullet Lacks advanced citation metrics\\ \textbullet Limited interoperability with other tools\\ \textbullet Steep learning curve for beginners\end{tabular} \\ \hline
\multicolumn{1}{|l|}{6}            & \multicolumn{1}{l|}{\begin{tabular}[c]{@{}l@{}}Publish or Perish\\ \citep{harzing2010publish}\end{tabular}
} & \multicolumn{1}{l|}{Software}                                                 & \multicolumn{1}{l|}{\begin{tabular}[c]{@{}l@{}}\textbullet h-index calculation\\ \textbullet Citation trend analysis\\ \textbullet Google Scholar\\ integration\end{tabular}}                 & \begin{tabular}[c]{@{}l@{}}\textbullet Dependent on Google Scholar\\ \textbullet Inconsistent citation data\\ \textbullet Lacks real-time monitoring\end{tabular}                            \\ \hline
\multicolumn{1}{|l|}{7}            & \multicolumn{1}{l|}{\begin{tabular}[c]{@{}l@{}}SciMAT\\ \citep{cobo2012scimat}\end{tabular}
}                                    & \multicolumn{1}{l|}{Software}                                                 & \multicolumn{1}{l|}{\begin{tabular}[c]{@{}l@{}}\textbullet Co-word analysis\\ \textbullet Performance analysis\\ \textbullet Science mapping\\ \textbullet Trend analysis\end{tabular}}                   & \begin{tabular}[c]{@{}l@{}}\textbullet Data preprocessing required\\ \textbullet Lacks interactive visualization\\ \textbullet Limited support for non-English texts\end{tabular}            \\ \hline
\multicolumn{1}{|l|}{8}            & \multicolumn{1}{l|}{\begin{tabular}[c]{@{}l@{}}CitNetExplorer\\ \citep{van2014citnetexplorer}\end{tabular}
}                            & \multicolumn{1}{l|}{Software}                                                 & \multicolumn{1}{l|}{\begin{tabular}[c]{@{}l@{}}\textbullet Citation relationship\\ mapping\\ \textbullet Paper clustering\\ \textbullet Timeline visualization\end{tabular}}                  & \begin{tabular}[c]{@{}l@{}}\textbullet Limited customization options\\ \textbullet Lacks real-time citation updates\\ \textbullet High memory usage for large datasets\end{tabular}          \\ \hline
\multicolumn{1}{|l|}{9}            & \multicolumn{1}{l|}{\begin{tabular}[c]{@{}l@{}}CiteSpace\\ \citep{chen2004searching}\end{tabular}
}                                 & \multicolumn{1}{l|}{Software}                                                 & \multicolumn{1}{l|}{\begin{tabular}[c]{@{}l@{}}\textbullet Citation burst detection\\ \textbullet Knowledge mapping\\ \textbullet Co-citation analysis\\ \textbullet Document clustering\end{tabular}}    & \begin{tabular}[c]{@{}l@{}}\textbullet Complex interface\\ \textbullet Requires high computational resources\\ \textbullet Limited data export options\end{tabular}                          \\ \hline
\multicolumn{1}{|l|}{10}           & \multicolumn{1}{l|}{\begin{tabular}[c]{@{}l@{}}Sci2 Tool\\ \citep{borner2011science}\end{tabular}
}                                 & \multicolumn{1}{l|}{Software}                                                 & \multicolumn{1}{l|}{\begin{tabular}[c]{@{}l@{}}\textbullet Scientific workflow\\ analysis\\ \textbullet Bibliometric tool\\ integration\end{tabular}}                             & \begin{tabular}[c]{@{}l@{}}\textbullet Dated interface\\ \textbullet Steep learning curve\\ \textbullet Lacks cloud-based deployment\end{tabular}                                            \\ \hline
\end{tabular}
\end{table}

\subsection{Comparative analysis of existing tools}
The landscape of bibliometric and scientometric analysis has evolved with the development of various tools, each offering distinct functionalities tailored to specific research needs. This comparative analysis evaluates prominent tools such as Biblioshiny, BibExcel, ScientoPy, pyBibX, VOSviewer, Publish or Perish, SciMAT, CitNet Explorer, CiteSpace, and Sci2 Tool—against the backdrop of the proposed web-based platform, MetaInfoSci.

Biblioshiny helps with bibliometric research by letting users do tasks such as analysing co-authorship and looking at networks of keywords used together. On the other hand, you can’t merge your databases with AI-derived information using it. BibExcel supports both data preprocessing and statistical analysis without difficulty, although visualizing the network and using AI still require you to do the work manually. ScientoPy centers on analyzing time frames of scientific work, although it is not equipped for network or AI analysis~\citep{yang2024biblioshiny}.

Distinguished by the addition of AI, PyBibX uses topic modeling and text summarization made possible by BERT and ChatGPT. You can study entire networks such as those for citations and collaboration, with assistance from the software and create your own special queries. Yet, because it is a Python library, users who do not program may find it hard to use~\citep{pereira2023pybibx}.

VOSviewer specializes in making and displaying bibliographic networks, providing understandable maps of co-authorship, co-citation and keywords appearing together. New findings suggest its usefulness in examining text data in addition to conventional bibliometric data. Even so, it doesn't have artificial intelligence features or sophisticated query functions~\citep{van2017citation, altay2023vosviewer}.

Publish or Perish lets you access citations from different databases, though it offers no support for social network research or AI. Scientific literature mapping and thematic review analysis are supported by SciMAT, but it doesn't allow for many custom queries or AI access. CitNet Explorer makes it easy to visualize citation networks at the publication level but it doesn't have full bibliometric capabilities or AI benefits. Although CiteSpace shows research fronts and rising trends by looking at citation bursts, the application needs training and doesn't use AI. Sci2 Tool permits a wide array of network analysis but is not able to produce custom visualizations or include AI features~\citep{moral2020software}.

In response, MetaInfoSci was made to overcome these challenges by building an integrated online platform that puts the best features of separate tools side by side. It allows data from many bibliographic databases to be combined which enables extensive analysis from a wide range of sources. You can create customized queries and visuals on the platform to focus on what matters to your study. Due to its AI integration, MetaInfoSci provides users with summary of bibliometric information which makes the data easier to interpret and decide upon. Because the user interface is simple, anyone can use it, allowing many researchers to work with advanced bibliometric data.

All in all, the tools that currently exist have useful features, but they usually don't link together and rely on AI. MetaInfoSci provides a complete solution by joining together data integration, flexible analyses and AI, thus helping advance bibliometric and scientometric research.

\section{Methodology}
\begin{figure}[!h]
    \centering
\includegraphics[width=0.6\linewidth]{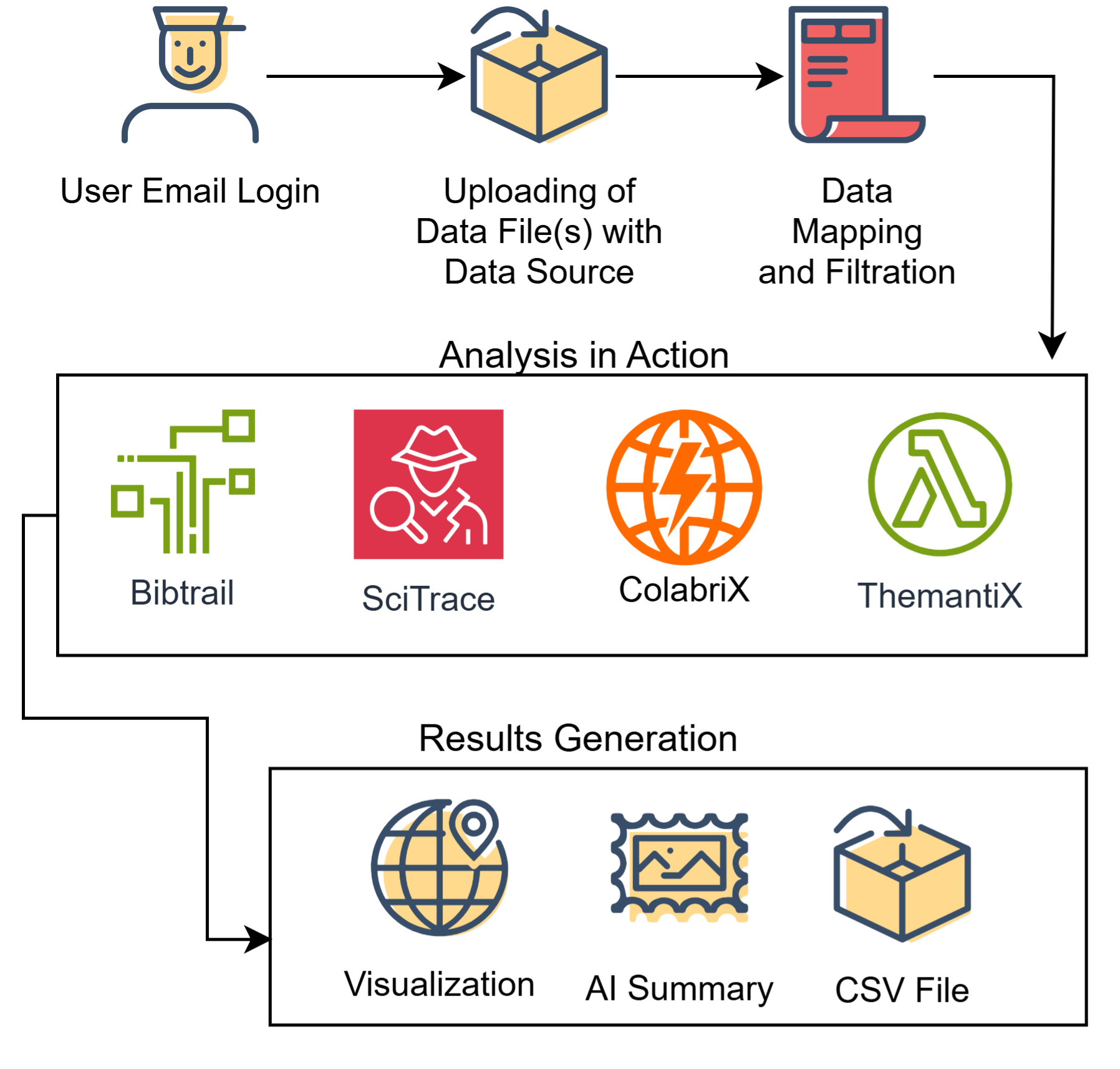} 
\caption{MetaInfoSci work flow.}
\label{fig:flowchart}
\end{figure}

The flowchart shown in Fig.~\ref{fig:flowchart} illustrates the workflow of a web-based platform for scholarly data analysis. The process begins with user authentication via email login, followed by the uploading of bibliographic data files from sources like Scopus or Web of Science. These files then undergo data mapping and filtration to ensure relevance and consistency. The cleaned data is analyzed using four core modules: Bibtrail for bibliometric analysis, SciTrace for tracking scientific impact, ColabriX for exploring collaboration networks, and ThemantiX for thematic or keyword-based analysis. The results from these modules are compiled into interactive visualizations, AI-generated textual summaries, and downloadable CSV files, offering users a comprehensive and customizable overview of their research data.

\subsection{Quartile analysis}
To perform quartile-related analysis, a master file containing journal quartile information was downloaded from Scimago(\url{https://www.scimagojr.com/}). This file was then mapped to the ISSN numbers of the journals provided in the dataset.
\subsection{Gender analysis}
 For gender identification, each author's name, along with their country information, was extracted from the dataset. Using AI-based prediction models, the likely gender of each author was then determined.

\subsection{Centrality measures}
Centrality measures are key metrics in network analysis used to identify the most important or influential nodes within a network based on their connectivity and position relative to other nodes. 
\subsubsection{Degree centrality}
It measures the number of nodes a certain node is connected to. This helps us infer the most connected node in the network and can be calculated as (Eq~\ref{eq:DC})

\begin{equation}
C_D(v) = \deg(v)
\label{eq:DC}
\end{equation}
where \( \deg(v) \) is the degree (number of edges) of node \(v\).
\subsubsection{Betweenness centrality}
This quantifies the number of times a node acted as a bridge in the shortest path between two nodes. It is crucial for understanding which nodes are responsible for controlling the information flow between nodes and can be calculated as (Eq~\ref{eq:BC})

\begin{equation}
C_B(v) = \sum_{s \neq v \neq t} \frac{\sigma_{st}(v)}{\sigma_{st}}
\label{eq:BC}
\end{equation}
where \( \sigma_{st} \) is the total number of shortest paths from node \(s\) to node \(t\) and \( \sigma_{st}(v) \) is the number of shortest paths from \(s\) to \(t\) that pass through node \(v\).
\subsubsection{Closeness centrality}
It shows how close a node is to all other nodes on the network, based on shortest path. It is reciprocal of the sum of shortest path to all the nodes and can be calculated as (Eq~\ref{eq:CoC})

\begin{equation}
C_C(v) = \frac{1}{\sum_{t} d(v,t)}
\label{eq:CoC}
\end{equation}
where \(d(v,t)\) is the shortest distance between node \(v\) and node \(t\).
\subsubsection{Eigenvector centrality}
Eigenvector centrality measures a node's influence in a network by assigning higher scores to nodes that are connected to other well-connected (high-scoring) nodes. It can be calculated as (Eq~\ref{eq:EC})

\begin{equation}
    C_E(v) = \frac{1}{\lambda} \sum_{u \in N(v)} A_{vu} C_E(u)
    \label{eq:EC}
\end{equation}
where \( \lambda \) is a constant (eigenvalue), \( A_{vu} \) is the adjacency matrix entry between nodes \(v\) and \(u\), \( C_E(u) \) is the eigenvector centrality of node \(u\), and \( N(v) \) is the set of neighbors of node \(v\).

\section{Core Features}
\subsection{Multi-file upload \& automated preprocessing}
\begin{itemize}
    \item Accepts multiple data files from diverse bibliographic databases.
    \item Automatically merges, deduplicates, and maps fields to the required internal format.
    \item If necessary, users can manually adjust field mappings via an intuitive interface.
\end{itemize}

\begin{figure}[htbp]
    \centering
\includegraphics[width=0.7\linewidth]{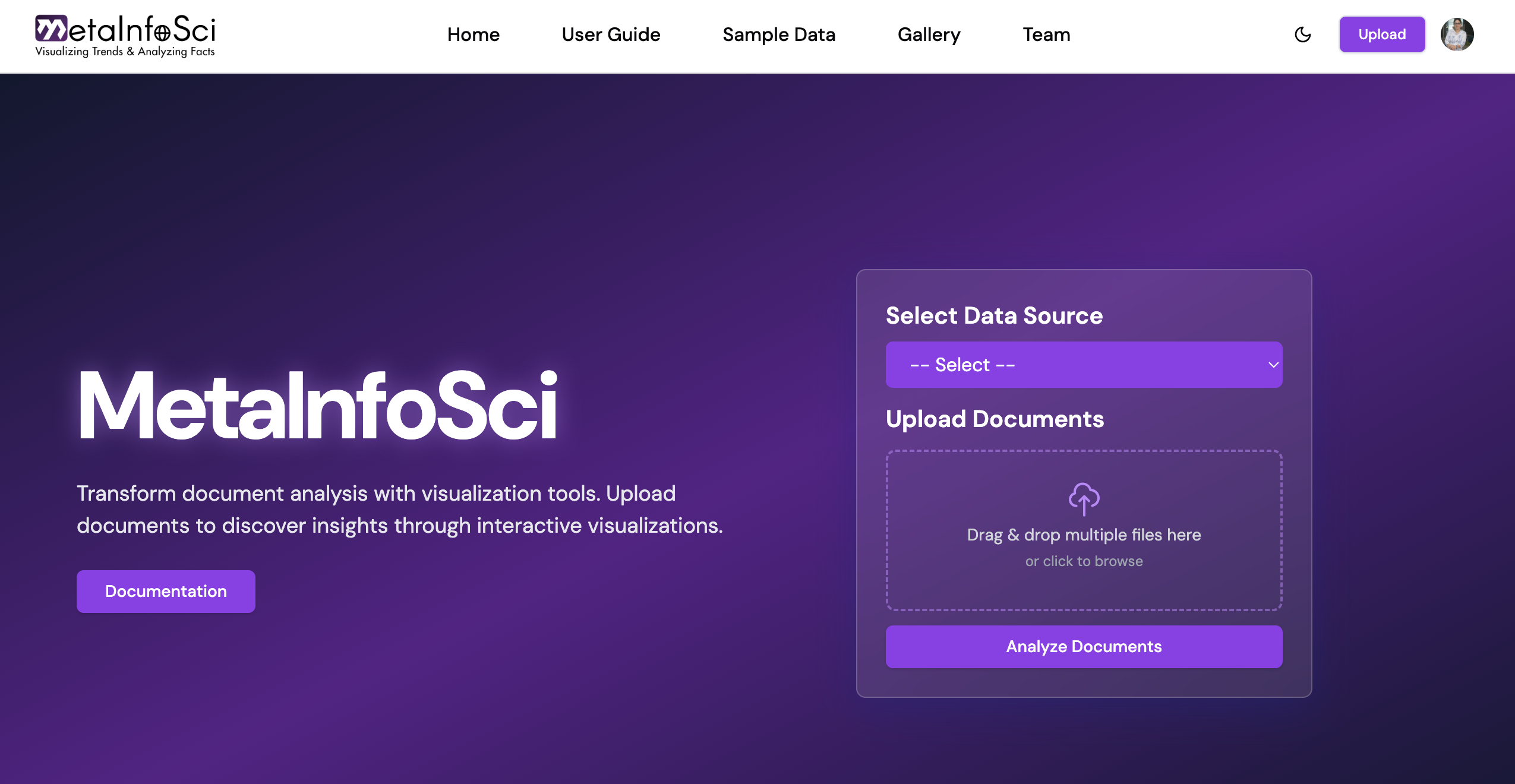} 
\caption{Upload file page.}
\label{fig:upload}
\end{figure}

\begin{figure}[htbp]
    \centering
\includegraphics[width=0.7\linewidth]{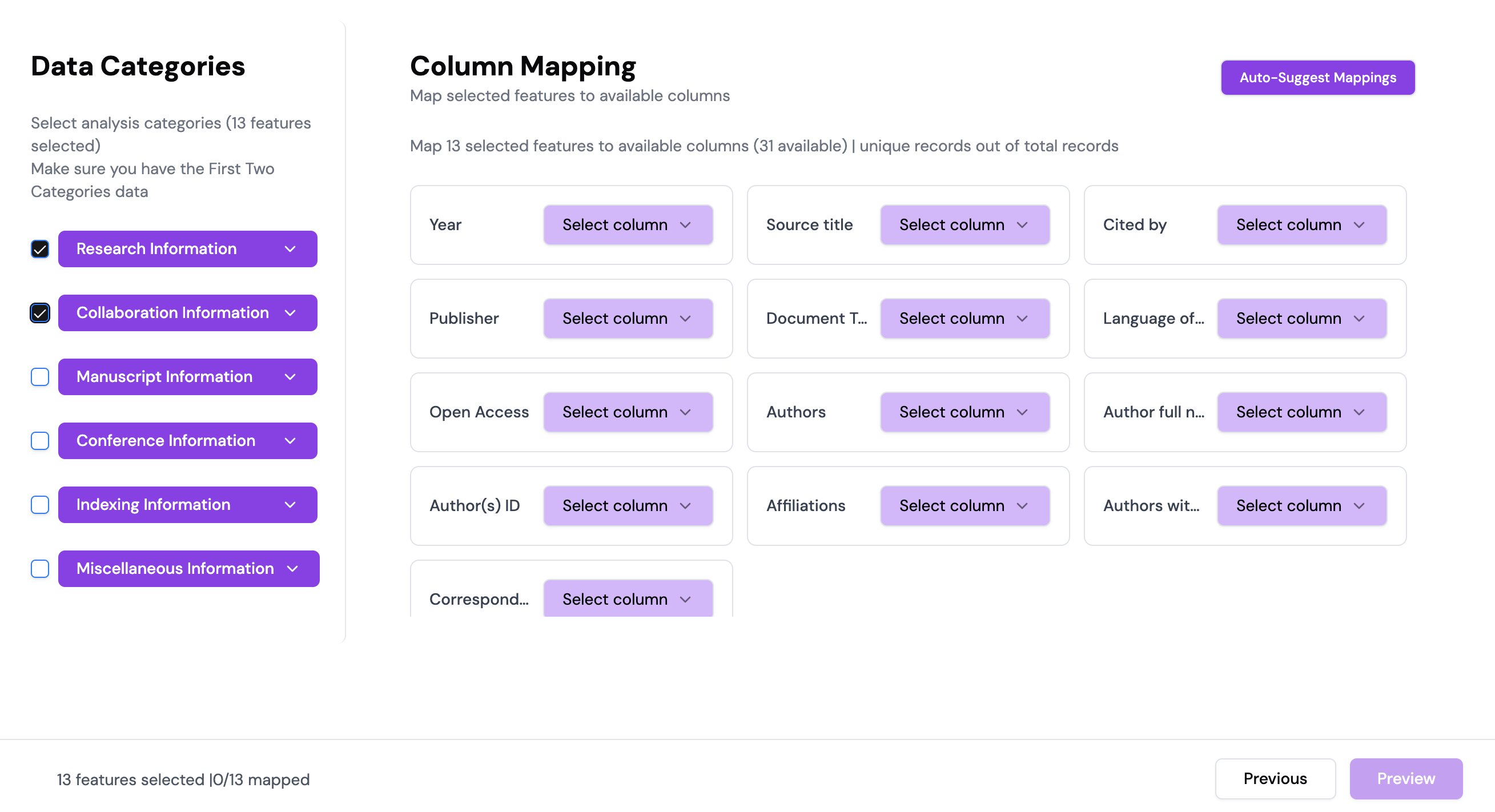} 
\caption{Mapping data fields to required format.}
\label{fig:mapping}
\end{figure}

\begin{figure}[htbp]
    \centering
\includegraphics[width=0.7\linewidth]{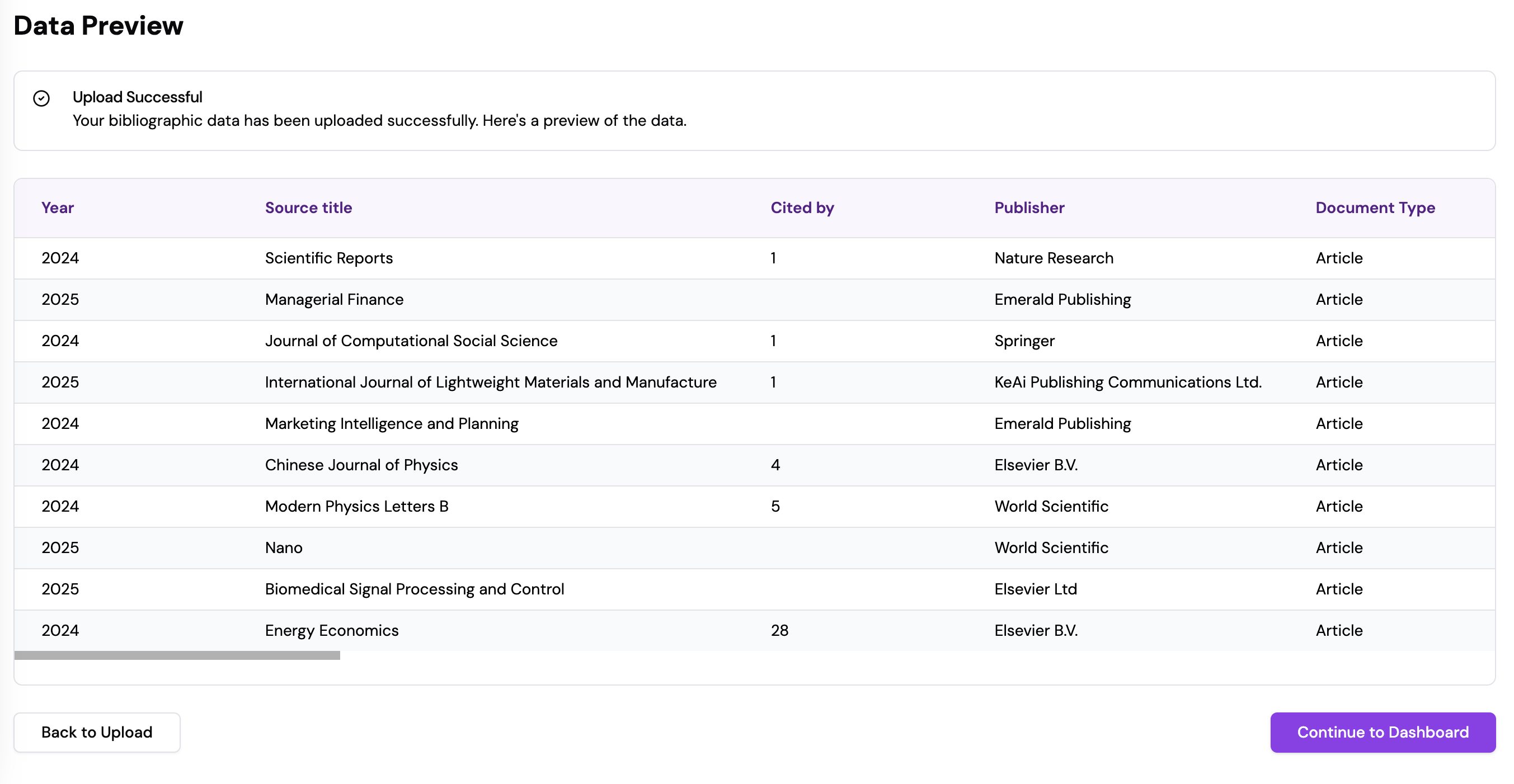} 
\caption{After mapping top 10 data entries are displayed.}
\label{fig:preview}
\end{figure}


\subsection{Data overview \& custom filtering}
\begin{itemize}
    \item Displays the uploaded dataset with a summary of key statistics (e.g., number of records, authors, institutions, countries).
    \item Offers filtering tools to refine datasets based on user-defined criteria before performing in-depth analyses.
\end{itemize}

\begin{figure}[htbp]
    \centering
\includegraphics[width=0.7\linewidth]{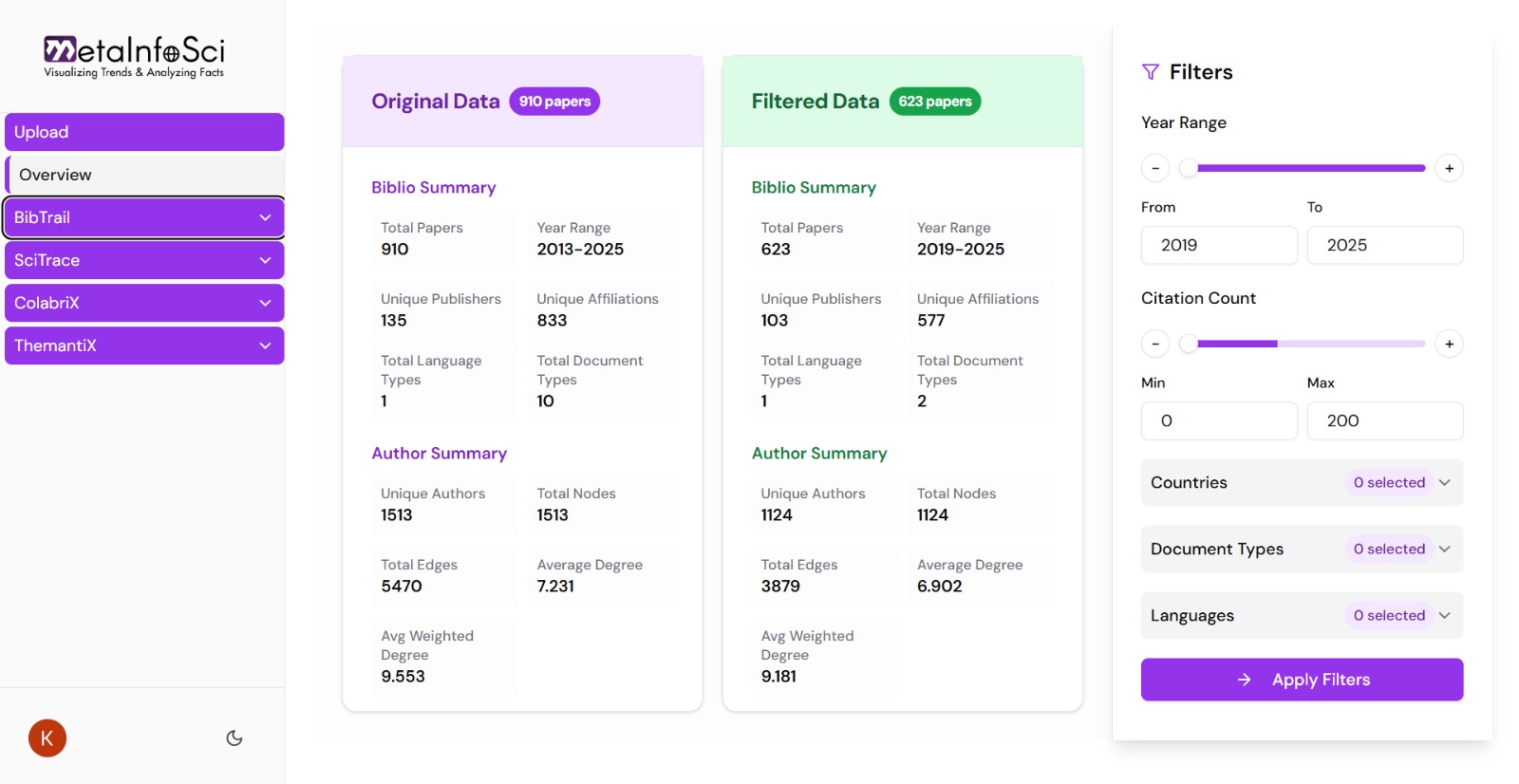} 
\caption{After mapping top 10 data entries are displayed.}
\label{fig:overview}
\end{figure}

\subsection{Comprehensive analytical capabilities}
Supports a wide range of analyses including:
\begin{itemize}
    \item Bibliometric analysis (e.g., citations, publications, journal impact)
    \item Scientometric analysis (e.g., research trends, gender analysis, Teamsize, authorship)
    \item Collaboration and network analysis (e.g., co-authorship, institutional networks, country-level collaborations)
    \item Thematic analysis (e.g., thematic evolution, keywords frequency)

\end{itemize}

\subsubsection{BibTrail: Bibliometric analysis}
\begin{figure}[htbp]
    \centering
\includegraphics[width=0.7\linewidth]{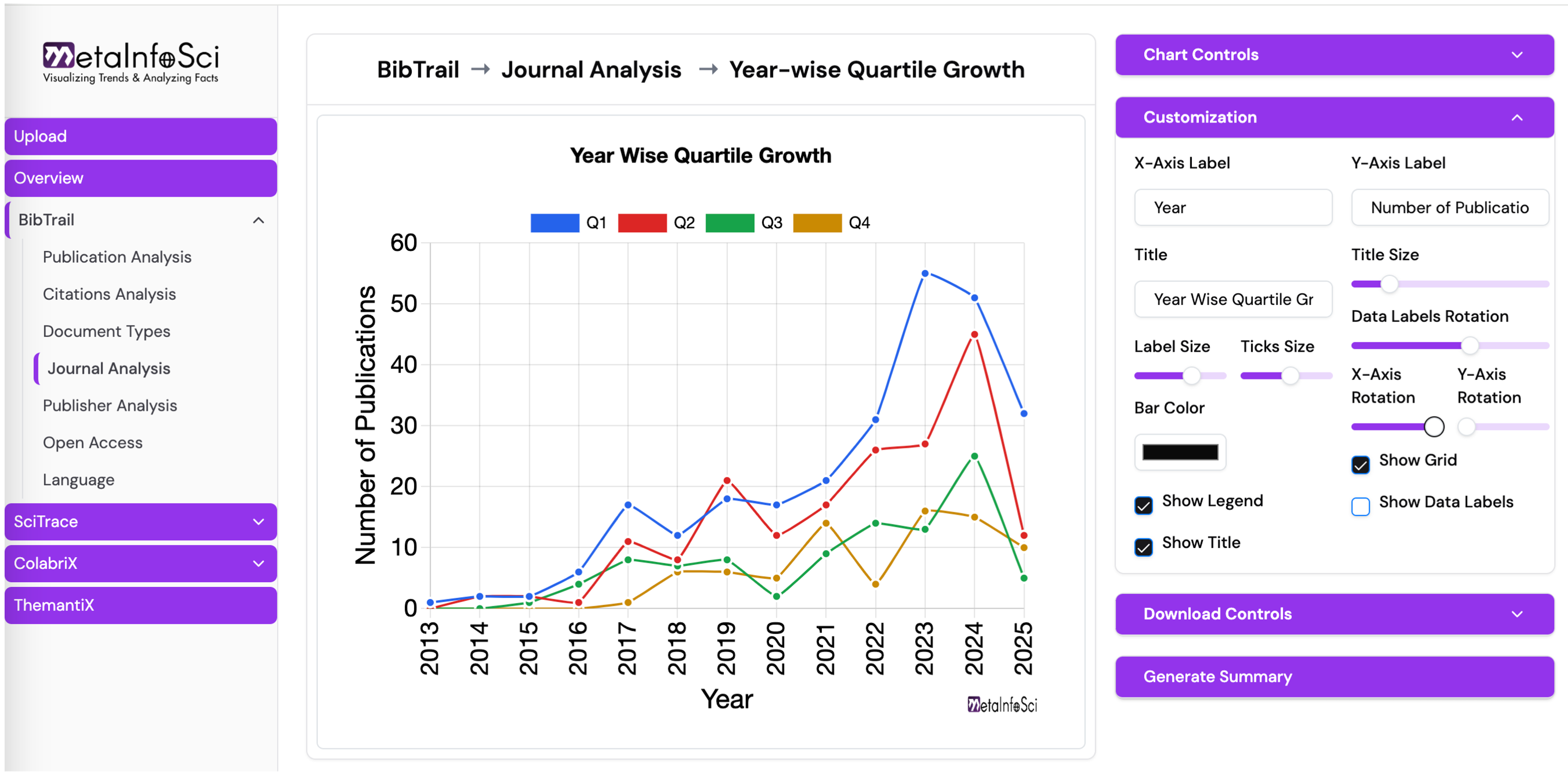} 
\caption{Upload file page.}
\label{fig:BibTrail}
\end{figure}

This section demonstrates the bibliometric analysis (see Fig~\ref{fig:BibTrail}) . Table~\ref{Tab:BibTrail} highlights the key functions in it.

\begin{table}[htbp]
\caption{List of BibTrail features.}
\label{Tab:BibTrail}
\begin{tabular}{|l|l|l|l|}
\hline
\textbf{Functions}                                             & \textbf{Analysis Types}    & \textbf{Description} & \textbf{Plot Type}    \\ \hline
\begin{tabular}[c]{@{}l@{}}Publication\\ Analysis\end{tabular} & \begin{tabular}[c]{@{}l@{}}\textbullet Total Count\\ \textbullet Cumulative\\ \textbullet Proportion\end{tabular}                                                                                       & \begin{tabular}[c]{@{}l@{}}Calculate the total number\\ of publications per year,\\ cumulative growth and\\ year wise proportion of\\ papers published\end{tabular}                                                                                                                                 & \begin{tabular}[c]{@{}l@{}}\textbullet Bar chart \\ \textbullet Line plot\end{tabular}                                                 \\ \hline
\begin{tabular}[c]{@{}l@{}}Citations\\ Analysis\end{tabular}   & \begin{tabular}[c]{@{}l@{}}\textbullet Total Count\\ \textbullet Average\\ \textbullet Cumulative\\ \textbullet Proportion\\ \textbullet Yearwise Citations\end{tabular}                                                        & \begin{tabular}[c]{@{}l@{}}Calculate the total number\\ of citations received per year,\\ average number of citations\\ received per year, cumulative\\ growth, year wise proportion\\ of citations received, and\\ year wise citations distribution\end{tabular}                                   & \begin{tabular}[c]{@{}l@{}}\textbullet Bar chart \\ \textbullet Line plot\\ \textbullet Box Plot\\ \textbullet Violin Plot\\ \textbullet Swarm Plot\end{tabular}           \\ \hline
\begin{tabular}[c]{@{}l@{}}Document\\ Types\end{tabular}       & \begin{tabular}[c]{@{}l@{}}\textbullet Total Count\\ \textbullet Yearwise Count\\ \textbullet Decade wise\\ \textbullet Documents Vs Citations\end{tabular}                                                          & \begin{tabular}[c]{@{}l@{}}Calculate the total number\\ of publications in each\\ document category, year\\ wise and decade wise\\ count against each type,\\ and for each document\\ type compute the\\ distribution of citations\end{tabular}                                                     & \begin{tabular}[c]{@{}l@{}}\textbullet Bar Chart\\ \textbullet Pie Chart\\ \textbullet Doughnut\\ \textbullet Box Plot\\ \textbullet Swarm Plot\end{tabular}               \\ \hline
\begin{tabular}[c]{@{}l@{}}Journal\\ Analysis\end{tabular}     & \begin{tabular}[c]{@{}l@{}}\textbullet Total Count\\ \textbullet Quartile Count\\ \textbullet Quartile wise top Journals\\ \textbullet Year wise growth in each Quartile\\ \textbullet Top journals in each quartile\end{tabular} & \begin{tabular}[c]{@{}l@{}}Calculate the total number\\ of publications in each\\ listed journal and display\\ top among them. Map \\ Journal name with\\ Scimago Journal\\ Quartile ranking and\\ display the count in\\ each quartile, Quartile\\ wise journal list and\\ its growth\end{tabular} & \begin{tabular}[c]{@{}l@{}}\textbullet Bar Chart\\ \textbullet Pie Chart\\ \textbullet Line Plot\\ \textbullet Box Plot\\ \textbullet Swarm Plot\\ \textbullet Stack Plot\end{tabular} \\ \hline
\begin{tabular}[c]{@{}l@{}}Publisher\\ Analysis\end{tabular}   & \begin{tabular}[c]{@{}l@{}}\textbullet Total Count\\ \textbullet Number of journals per publisher\end{tabular}     & \begin{tabular}[c]{@{}l@{}}Calculate the total number\\ of publications in each\\ listed publisher and display\\ top among them\end{tabular}                                                                                                                                                        & \begin{tabular}[c]{@{}l@{}}\textbullet Bar Chart\\ \textbullet Pie Chart\\ \textbullet Doughnut\end{tabular}                                       \\ \hline
\begin{tabular}[c]{@{}l@{}}Open\\ Access\end{tabular}          & \begin{tabular}[c]{@{}l@{}}\textbullet Total Count\\ \textbullet Open-Access Vs Citations\end{tabular}                                                                                       & \begin{tabular}[c]{@{}l@{}}Calculate the total number\\ of publications in each\\ category,  and for each\\ open-access type compute\\ the distribution of citations\end{tabular}                                                                                                                   & \begin{tabular}[c]{@{}l@{}}\textbullet Bar Chart\\ \textbullet Pie Chart\\ \textbullet Doughnut\\ \textbullet Box Plot\\ \textbullet Swarm Plot\end{tabular}               \\ \hline
Language                                                       & \textbullet Total Count                                                                                                                                                          & \begin{tabular}[c]{@{}l@{}}Calculate the total number\\ of publications in each\\ language category\end{tabular}                                                                                & \begin{tabular}[c]{@{}l@{}}\textbullet Bar Chart\\ \textbullet Pie Chart\\ \textbullet 
 Doughnut\end{tabular}                                       \\ \hline
\end{tabular}
\end{table}

\subsubsection{SciTrace: Scientometric analysis}
\begin{figure}[htbp]
    \centering
\includegraphics[width=0.7\linewidth]{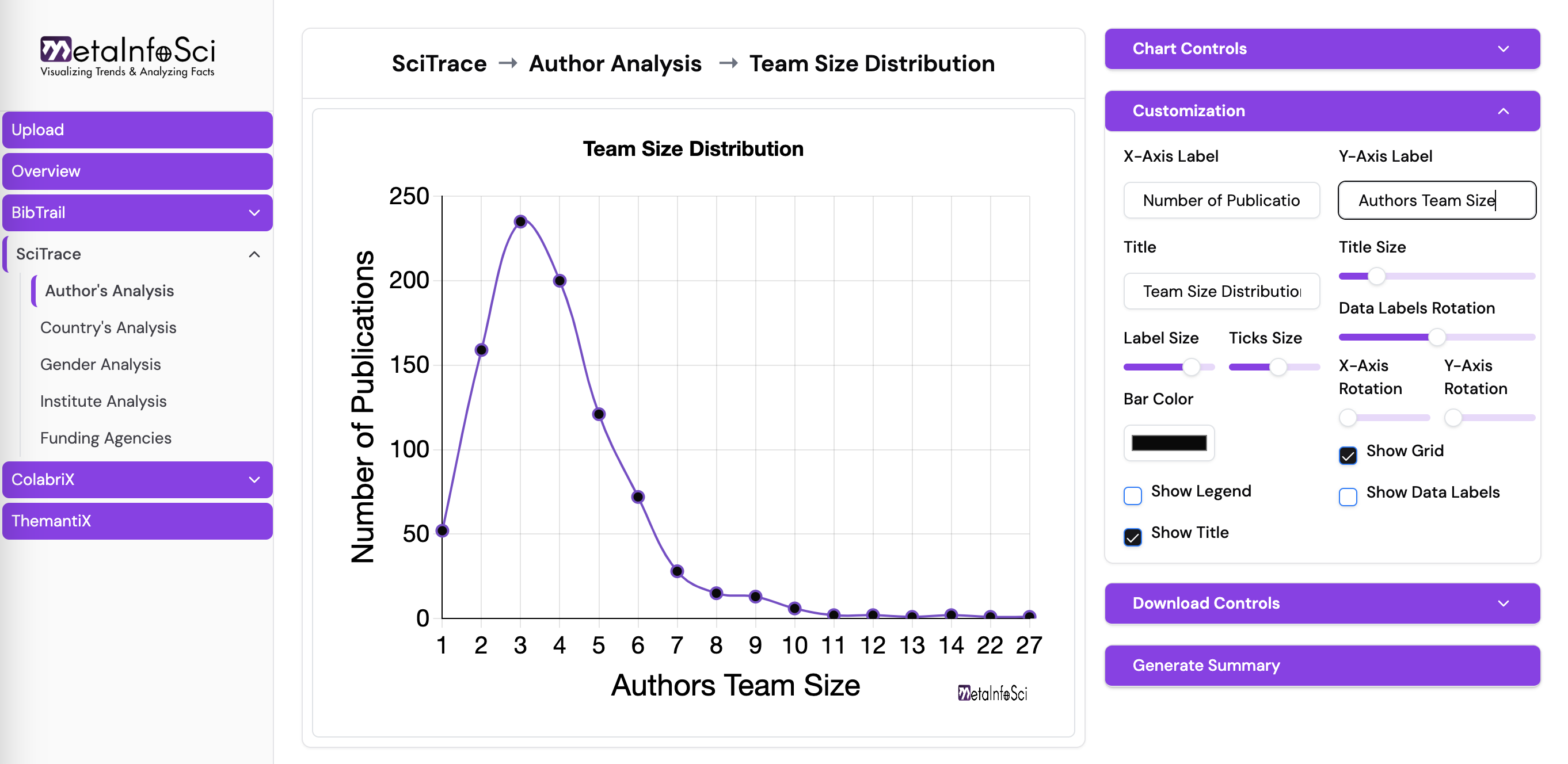} 
\caption{Upload file page.}
\label{fig:SciTrace}
\end{figure}

This section demonstrates the scientometric analysis (see Fig~\ref{fig:SciTrace}). Table~\ref{Tab:SciTrace} highlights the key functions in it.

\begin{table}[]
\caption{List of SciTrace features.}
\label{Tab:SciTrace}
\begin{tabular}{|l|l|l|l|}
\hline
\textbf{SciTrace   Functions} & \textbf{Analysis Types}  & \textbf{Description}  & \textbf{Plot Type}                                                           \\ \hline
Author's Analysis             & \begin{tabular}[c]{@{}l@{}}\textbullet Author Count\\ \textbullet Top Author's List\\ \textbullet Team Size\\ \textbullet Collaboration\end{tabular}                                       & \begin{tabular}[c]{@{}l@{}}Number of papers published\\ by number of authors, highlight\\ authors based on number of\\ publications, size of collaboration\\ (Team size) in each paper,\\ and how frequent people\\ collaborate\end{tabular}                               & \begin{tabular}[c]{@{}l@{}}\textbullet Bar Chart\\ \textbullet Scatter Plot\end{tabular}             \\ \hline
Country's Analysis            & \begin{tabular}[c]{@{}l@{}}\textbullet Country Count\\ \textbullet Lead Country Count\\ \textbullet Team Size\\ \textbullet Collaboration\\ \textbullet Papers Vs. Citations each country\end{tabular} & \begin{tabular}[c]{@{}l@{}}Number of papers published\\ Vs Citations received,\\ highlight countries based on\\ number of publications,\\ who is top as leader,\\ size of collaboration (Team size)\\ in each paper, and how\\ frequent countries collaborate\end{tabular} & \begin{tabular}[c]{@{}l@{}}\textbullet Bar Chart\\ \textbullet Scatter Plot\\ \textbullet World Map\end{tabular} \\ \hline
Gender Analysis               & \begin{tabular}[c]{@{}l@{}}\textbullet Total Count\\ \textbullet Authorship Position \\ \textbullet Country wise Gender Proportion\end{tabular}                                & \begin{tabular}[c]{@{}l@{}}Proportion of each gender,\\ and at which position gender\\ contributed. Display Top 10\\ countries male female\\ proportion\end{tabular}                                                                                                       & \begin{tabular}[c]{@{}l@{}}\textbullet Bar Chart\\ \textbullet Pie Chart\end{tabular}                \\ \hline
Top Institutes                & \textbullet Total Count                                                                                                                                & \begin{tabular}[c]{@{}l@{}}Highlight Institutes based\\ on number of publications\end{tabular}                                                                                                                                                                             & \textbullet Bar Chart                                                                    \\ \hline
Top Funding Agencies          & \textbullet Total Count                                                                                                                                & \begin{tabular}[c]{@{}l@{}}Highlight funding bodies\\ based on acknowledgement\\ in paper\end{tabular}                                                                                                                                                                     & \textbullet Bar Chart                                                                    \\ \hline
\end{tabular}
\end{table}

\subsubsection{ColabriX: Collaboration and network analysis}
\begin{figure}[htbp]
    \centering
\includegraphics[width=0.7\linewidth]{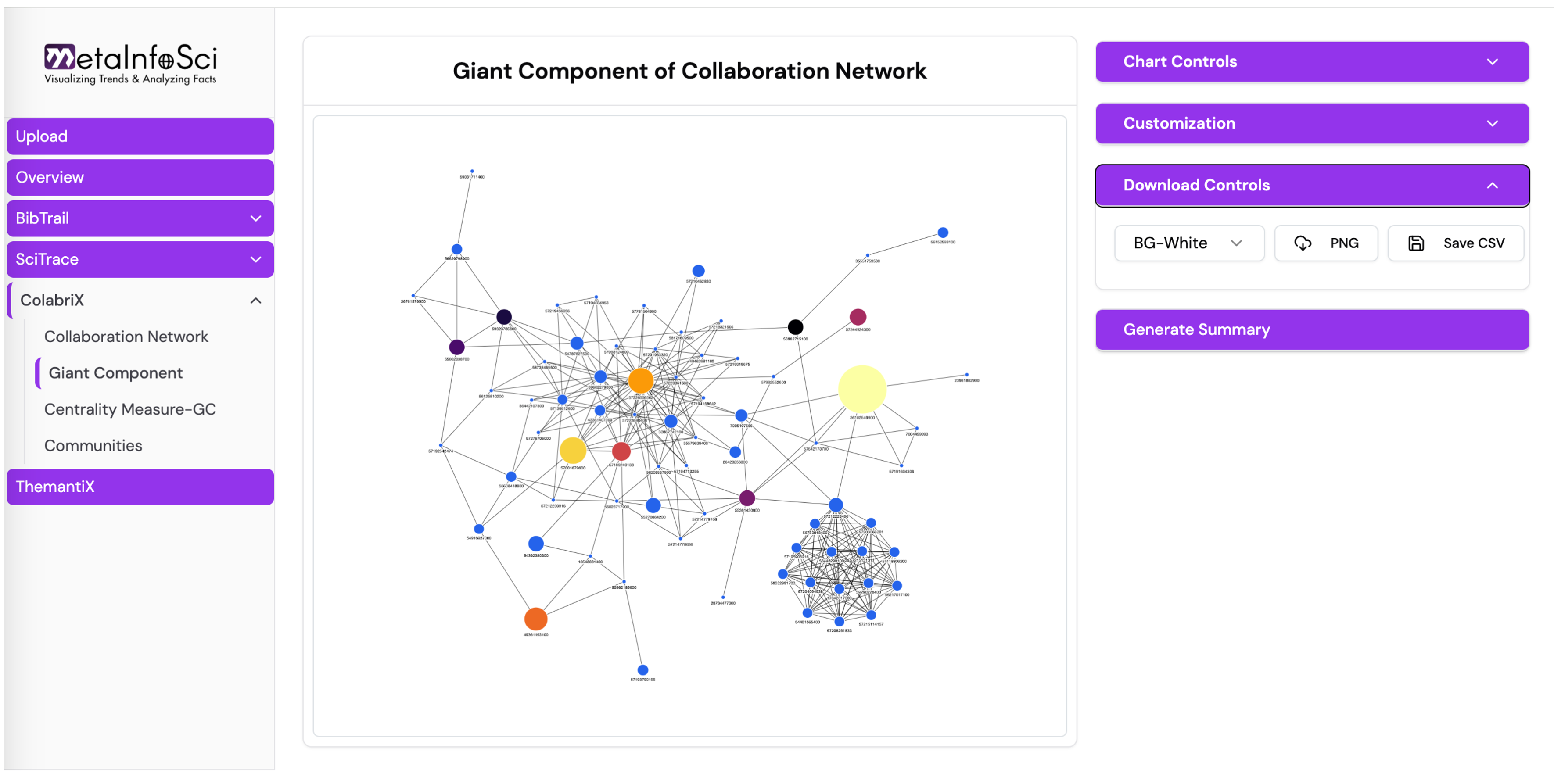} 
\caption{Upload file page.}
\label{fig:ColabriX}
\end{figure}

This section demonstrates the Collaboration and network analysis (see Fig~\ref{fig:ColabriX}). Table~\ref{Tab:ColabriX} highlights the key functions in it.

\begin{table}[htbp]
\caption{List of ColabriX features.}
\label{Tab:ColabriX}
\begin{tabular}{|l|l|l|l|l|l|}
\hline
\textbf{Functions}   & \textbf{Analysis Types}    & \textbf{Check-1}   & \textbf{Check-2} & \textbf{Description}& \textbf{Plot Type}  \\ \hline
\begin{tabular}[c]{@{}l@{}}Collaboration\\ Network\end{tabular} & Complete Network  & \multicolumn{1}{c|}{\textbf{-}}                                & \begin{tabular}[c]{@{}l@{}}\textbullet Author\\ \textbullet Country\end{tabular} & \begin{tabular}[c]{@{}l@{}}Display authors or country\\ collaboration network\end{tabular}  & \textbullet Graph          \\ \hline
\begin{tabular}[c]{@{}l@{}}Giant \\ Component\end{tabular}      & \begin{tabular}[c]{@{}l@{}}\textbullet 1st GC\\ \textbullet 2nd GC\end{tabular}   & \begin{tabular}[c]{@{}l@{}}\textbullet Network\\ \textbullet Distribution\end{tabular} & \begin{tabular}[c]{@{}l@{}}\textbullet Author\\ \textbullet Country\end{tabular} & \begin{tabular}[c]{@{}l@{}}Display the largest connected\\ component (sub-network) from\\ authors or country collaboration\\ network\end{tabular}    & \begin{tabular}[c]{@{}l@{}}\textbullet Graph\\ \textbullet Scatter Plot\end{tabular} \\ \hline
\begin{tabular}[c]{@{}l@{}}Centrality\\ Measure-GC\end{tabular} & \begin{tabular}[c]{@{}l@{}}\textbullet Degree\\ \textbullet Betweenness\\ \textbullet Closeness\\ \textbullet Eigenvector\end{tabular} & \begin{tabular}[c]{@{}l@{}}\textbullet Network\\ \textbullet Distribution\end{tabular} & \begin{tabular}[c]{@{}l@{}}\textbullet Author\\ \textbullet Country\end{tabular} & \begin{tabular}[c]{@{}l@{}}To find most prominent node\\ (author or country) in authors\\ or country collaboration network.\\ Highlight prominent nodes in\\ network\end{tabular} & \begin{tabular}[c]{@{}l@{}}\textbullet Graph\\ \textbullet Scatter Plot\end{tabular} \\ \hline
Communities  & \begin{tabular}[c]{@{}l@{}}\textbullet Girven-Newman\\ \textbullet Modularity\\ \textbullet Liden\end{tabular}             & \begin{tabular}[c]{@{}l@{}}\textbullet Network\\ \textbullet Distribution\end{tabular} & \begin{tabular}[c]{@{}l@{}}\textbullet Author\\ \textbullet Country\end{tabular} & \begin{tabular}[c]{@{}l@{}}To find communities based\\ on similarities in network\end{tabular}  & \begin{tabular}[c]{@{}l@{}}\textbullet Graph\\ \textbullet Scatter Plot\end{tabular} \\ \hline
\end{tabular}
\end{table}

\subsubsection{ThemantiX: Thematic analysis}
\begin{figure}[htbp]
    \centering
\includegraphics[width=0.7\linewidth]{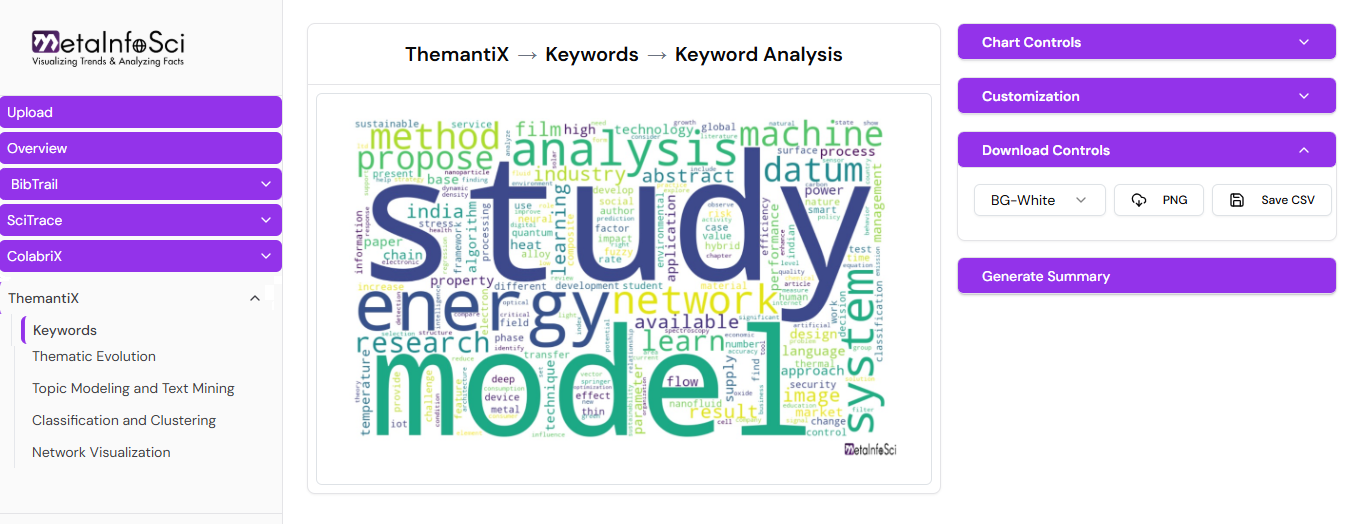} 
\caption{Upload data file page.}
\label{fig:ThemantiX}
\end{figure}

This section demonstrates the thematic analysis (see Fig~\ref{fig:ThemantiX}). Table~\ref{Tab:ThemantiX} highlights the key functions in it.

\begin{table}[htbp]
\caption{List of ThemantiX features.}
\label{Tab:ThemantiX}
\begin{tabular}{|l|l|l|l|}
\hline
\textbf{Functions} & \textbf{Analysis Types} & \textbf{Description} & \textbf{Plot Type} \\ \hline
Keywords & \begin{tabular}[c]{@{}l@{}}\textbullet Keywords Analysis\\ \textbullet Keywords Mapping\\ \textbullet Co-Word and\\ Co-Occurrence Analysis\end{tabular} & \begin{tabular}[c]{@{}l@{}}Count the frequency of each\\ keyword and map keywords\\ with filed or country; mapping\\ the intellectual structure of a\\ field and uncovering emerging\\ interdisciplinary areas\end{tabular} & \begin{tabular}[c]{@{}l@{}}\textbullet Word Cloud\\ \textbullet Mapping Graph\end{tabular} \\ \hline
Thematic Evolution & -- & \begin{tabular}[c]{@{}l@{}}Tracking the appearance\\ and frequency of specific\\ terms over time in titles\\ and abstracts can reveal\\ how research topics evolve\end{tabular} & Network \\ \hline
\begin{tabular}[c]{@{}l@{}}Topic Modeling and\\ Text Mining\end{tabular} & -- & \begin{tabular}[c]{@{}l@{}}Advanced computational\\ techniques, such as Latent\\ Dirichlet Allocation (LDA),\\ can be applied to abstracts\\ and titles to identify\\ underlying topics within a\\ corpus of literature. This\\ helps in uncovering hidden\\ thematic structures and\\ understanding the distribution\\ of topics across documents\end{tabular} & -- \\ \hline
\begin{tabular}[c]{@{}l@{}}Classification and\\ Clustering\end{tabular} & -- & \begin{tabular}[c]{@{}l@{}}By examining the textual\\ content of titles, abstracts,\\ and keywords, documents\\ can be grouped into clusters\\ representing specific research\\ areas or methodologies. This\\ classification aids in organizing\\ large datasets and identifying\\ niche research domains\end{tabular} & Scatter Plot\\ \hline
Network Visualization & --& \begin{tabular}[c]{@{}l@{}}Constructing networks based\\ on keyword co-occurrence or\\ thematic similarity allows for\\ the visualization of relationships\\ between different research topics.\\ These visual maps can highlight\\ central themes, peripheral topics,\\ and the interconnectedness of\\ various research areas\end{tabular} & Graph \\ \hline
\end{tabular}
\end{table}

\subsection{Customizable visualization tools}
Users can select and customize chart types, layouts, colors, and metrics for visualizing analysis results. Visualizations are downloadable in both image (PNG) and data (CSV) formats (see Table~\ref{tab:Viz}).

\begin{figure}[htbp]
    \centering
\includegraphics[width=0.8\linewidth]{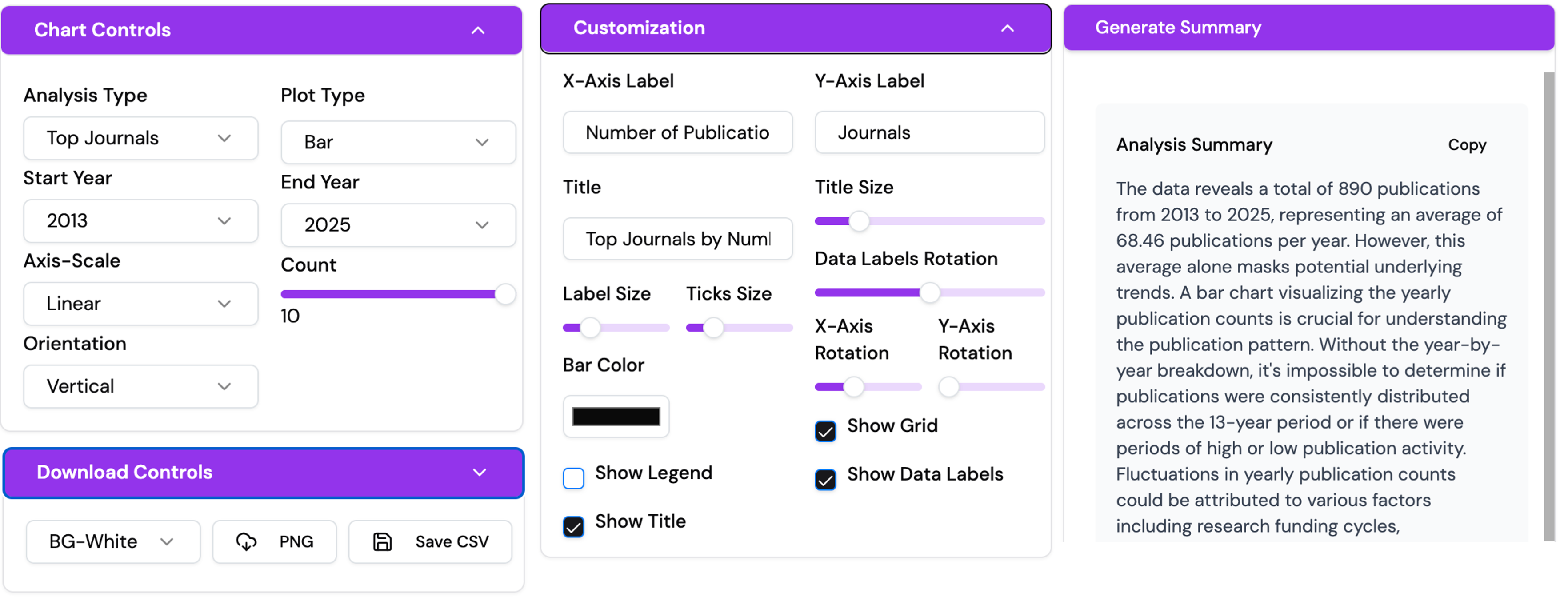} 
\caption{Plot customization includes (i) chart control panel, (ii) customization panel, and (iii) download options. (iv) generate AI summary.  }
\label{fig:Viz}
\end{figure}

\begin{enumerate}
    \item Chart control panel: Fig~\ref{fig:Viz}(left)) demonstrates the options given in chart control panel.

   \item Customization panel: Fig~\ref{fig:Viz}(middle)) demonstrates the options given in customization panel.

   \item Download panel: Fig~\ref{fig:Viz}(bottom)) demonstrates the options given in download panel.

   \item Generate AI-Summary: Each generated visualization is accompanied by an AI-generated explanatory summary, providing meaningful interpretation of the patterns and trends displayed (see Fig~\ref{fig:Viz}(right))).
\end{enumerate}

\begin{table}[htbp]
\caption{Plot customization options}
\label{tab:Viz2}
\begin{tabular}{|l|l|l|}
\hline
\textbf{Feature} & \textbf{Type} & \textbf{Description}  \\ \hline
\hline
\textbf{Plot customization options}  \\ \hline

Timeline         & Start year \& End Year & Select start and end year for analysis \\ \hline
Visualization    & \begin{tabular}[c]{@{}l@{}}\textbullet Bar Plot\\ \textbullet Line Plot\\ \textbullet Pie\\ \textbullet Doughnut\\ \textbullet Box Plot\end{tabular}                                           & Choose any plot layout for results     \\ \hline
Year Gap         & 1 to 5 Years & Display results with year gap          \\ \hline
Analysis         & \begin{tabular}[c]{@{}l@{}}\textbullet Citations: Total, Average, \\Cumulative, Median\\ \textbullet Publications: Total, Cumulative, \\Proportion\end{tabular} & Choose any analysis type               \\ \hline
Orientation      & \begin{tabular}[c]{@{}l@{}}\textbullet Vertical\\ \textbullet Horizontal\end{tabular}   & Orientation of bar plot                \\ \hline
Scale            & \begin{tabular}[c]{@{}l@{}}\textbullet Linear\\ \textbullet Log\end{tabular} & Choose X and Y Scale                   \\ \hline
Count            & Slider 1 to max  & In bar body, display top count         \\ \hline
Period           & \begin{tabular}[c]{@{}l@{}}\textbullet Year Wise\\ \textbullet Decade Wise\end{tabular}                                                                    & Choose analysis yearwise or decadewise  \\ \hline

\hline
\textbf{Plot customization options}                                                    \\ \hline
Adjust Labels    & \begin{tabular}[c]{@{}l@{}}\textbullet Name\\ \textbullet Fontsize\end{tabular}                                      & \begin{tabular}[c]{@{}l@{}}Adjust labels name and font size \\ for both X-Label and Y-Label \end{tabular} \\ \hline
Adjust Title     & \begin{tabular}[c]{@{}l@{}}\textbullet Name\\ \textbullet Fontsize\\ \textbullet Hide/View\end{tabular}                          & \begin{tabular}[c]{@{}l@{}}Adjust title name and fontsize \\ along with hide and view option \end{tabular}
 \\ \hline
Adjust Ticks     & \begin{tabular}[c]{@{}l@{}}\textbullet Fontsize\\ \textbullet Rotation\end{tabular}                                  & \begin{tabular}[c]{@{}l@{}}Adjust fontsize and rotation \\ of ticks for both X-axis and Y-axis\end{tabular}       \\ \hline
Color            & \begin{tabular}[c]{@{}l@{}}\textbullet Bar Color\\ \textbullet Border Color\\ \textbullet Line Color\\ \textbullet Marker Color\end{tabular} & \begin{tabular}[c]{@{}l@{}}Choose any color from color pallette for bars, \\ borders, lines and marker\end{tabular}
\\ \hline
Adjust Grid      & Hide/View  & Either display grid or hide                                             \\ \hline
Adjust Legend    & Hide/View                                                                                    & Either display legend or hide                                           \\ \hline
\hline
\textbf{Download Control Panel}         \\ \hline

Background       & \begin{tabular}[c]{@{}l@{}}\textbullet BG-W\\ \textbullet BG-T\end{tabular}                                          & \begin{tabular}[c]{@{}l@{}}To generate image with White (W) \\ and transparent (T) background\end{tabular}
 \\ 
\hline
Images           & \begin{tabular}[c]{@{}l@{}}\textbullet Png\\ \textbullet Jpeg\end{tabular}       & To generate png and jpeg format images \\ \hline
Data             & Csv                                                      & Download compiled data as csv file     \\ \hline
Share            & \begin{tabular}[c]{@{}l@{}}\textbullet Email\\ \textbullet WhatsApp\end{tabular} & Share image results                    \\ \hline
\end{tabular}
\label{tab:Viz}
\end{table}
\section{Results Demonstration}

To demonstrate the results, we used a sample dataset from Scopus for a university named\textit{ BML Munjal University}. A total of 890 records were filtered and downloaded, which were then uploaded to the MetaInfoSci portal for analysis.

\begin{figure}[htbp]
    \centering
\includegraphics[width=0.9\linewidth]{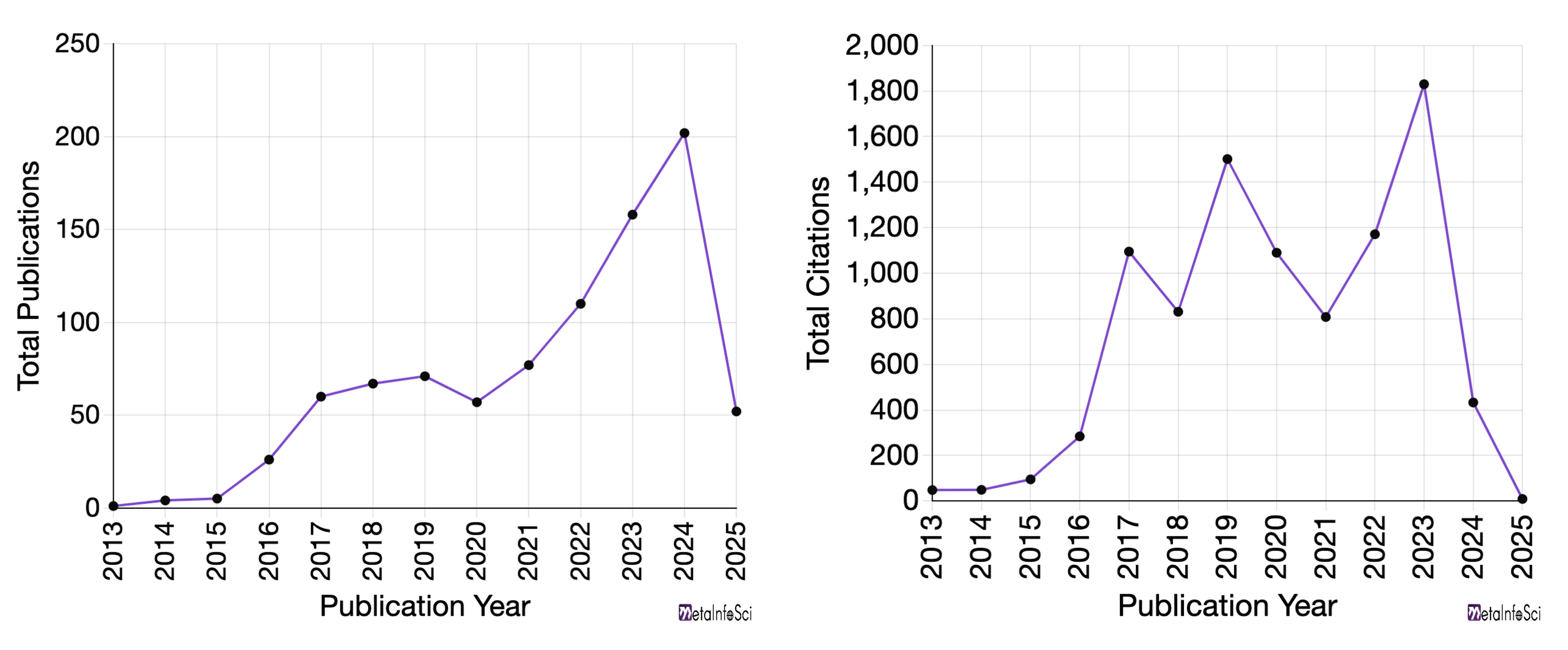} 
\caption{Growth and impact analysis: (\textit{left}) Total number of papers published per year. (\textit{right}) Total citations received per year.}
\label{fig:R1-PubYear}
\end{figure}
\begin{figure}[htbp]
    \centering
\includegraphics[width=0.9\linewidth]{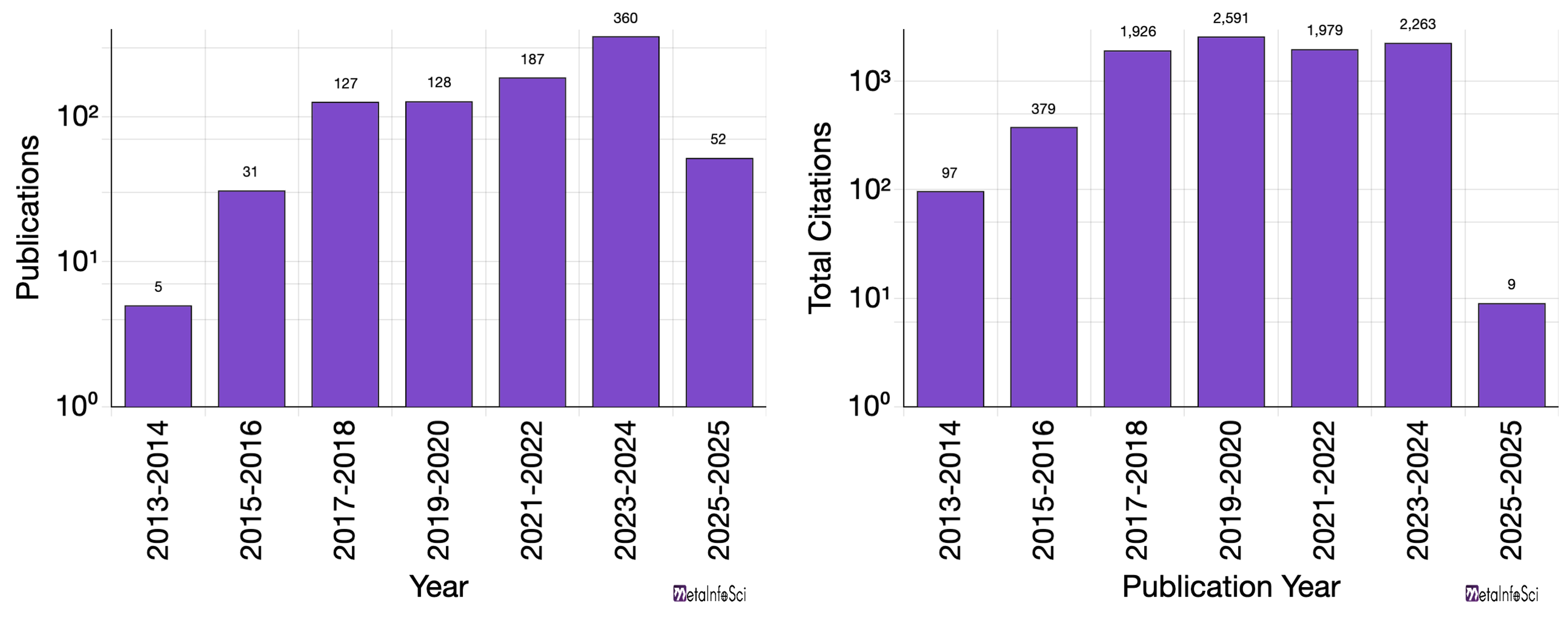} 
\caption{Growth and impact analysis: (\textit{left}) Total number of papers published in two years cycle. (\textit{right}) Total citations received in two years cycle.}
\label{fig:R1-2Y-PubYear}
\end{figure}

\begin{figure}[htbp]
    \centering
\includegraphics[width=0.9\linewidth]{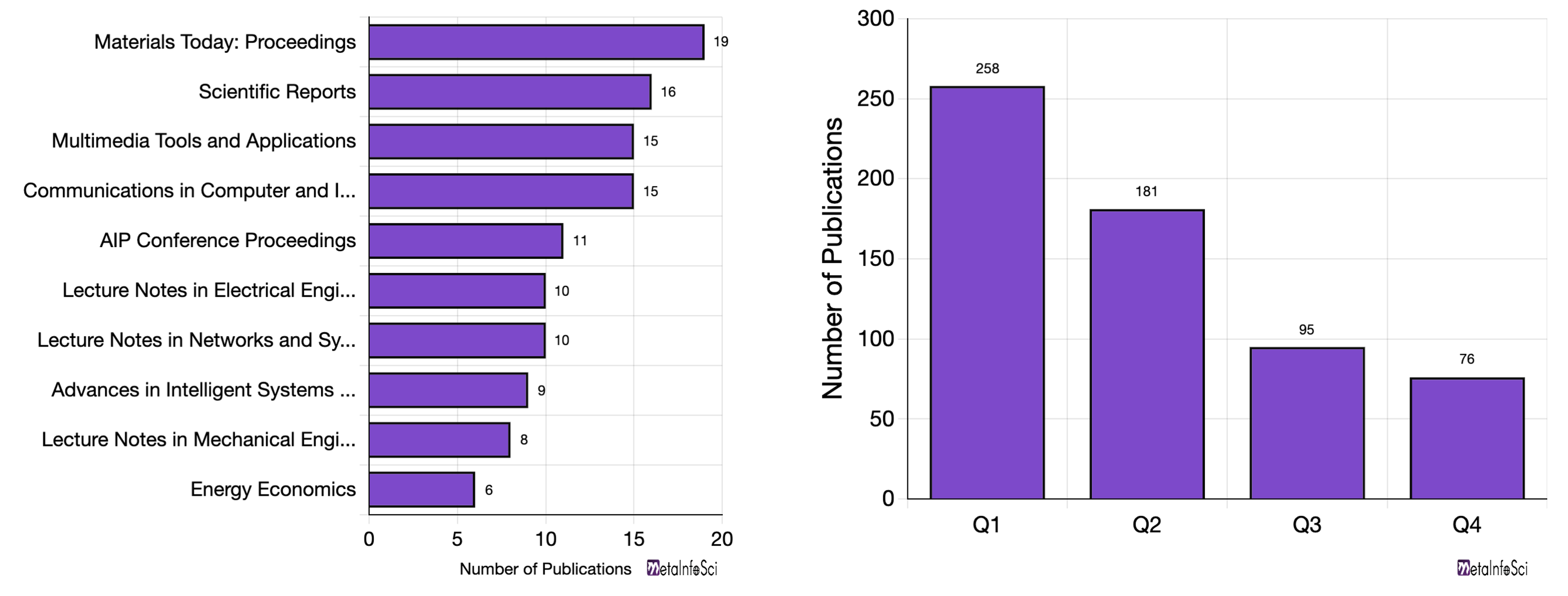} 
\caption{Journal and quartile analysis: (\textit{left}) Top 10 journals as per number of publications. (\textit{right}) Quartile wise paper distribution.}
\label{fig:R1-journal}
\end{figure}

\begin{figure}[htbp]
    \centering
    \includegraphics[width=0.9\linewidth]{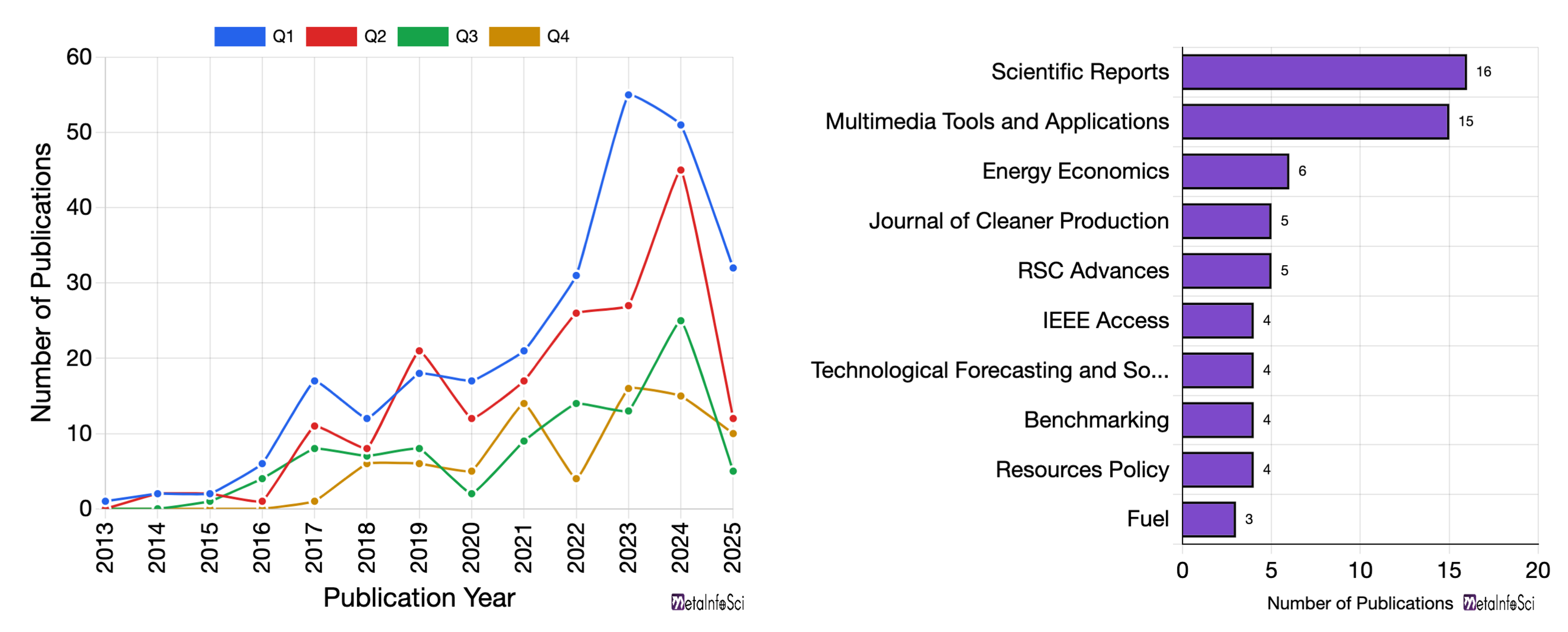} 
\caption{Journal and quartile analysis: (\textit{left}) Year wise paper distribution in each quartile. (\textit{right}) Top 10 journals as per number of publications in quartile one (Q1).  }
\label{fig:R1-Q1-Year}
\end{figure}

\begin{figure}[htbp]
    \centering
\includegraphics[width=0.9\linewidth]{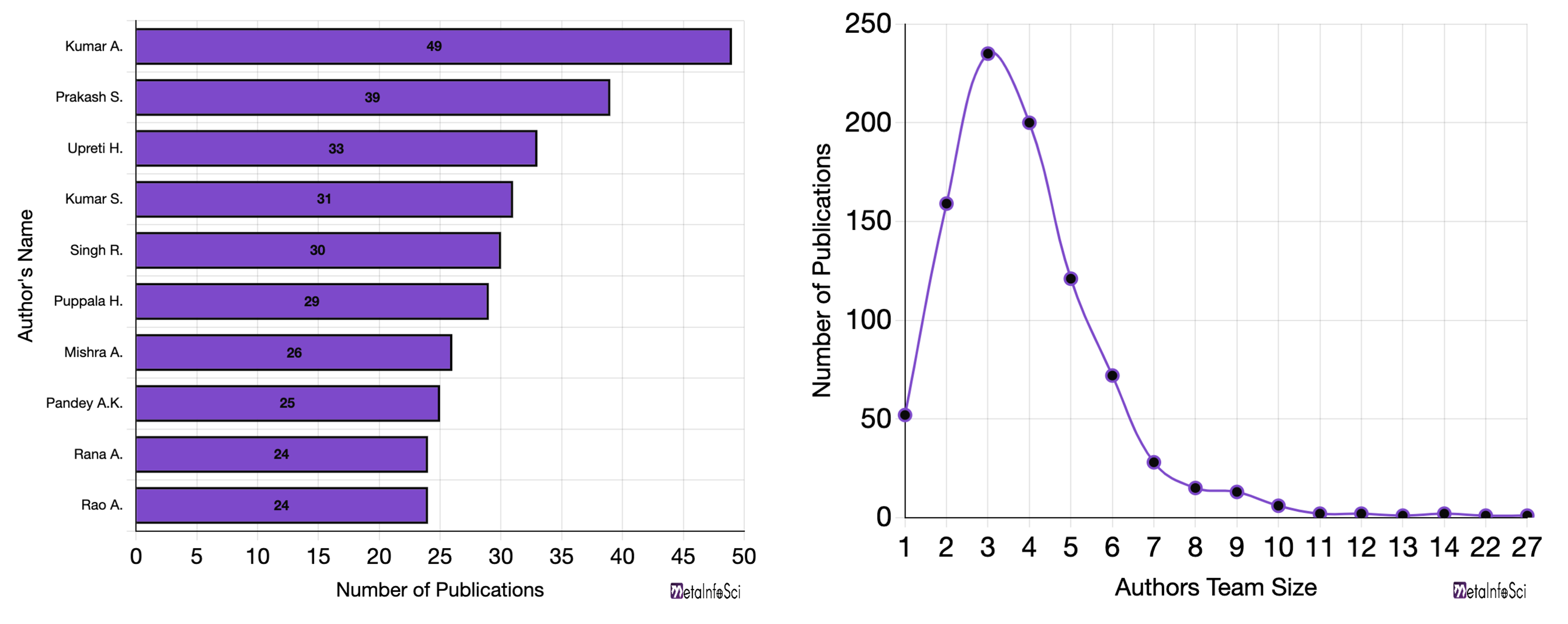} 
\caption{Author's analysis: (\textit{left}) Top 10 authors as per number of publications. (\textit{right}) Team wise paper distribution.}
\label{fig:R1-Authors}
\end{figure}

\begin{figure}[htbp]
    \centering
\includegraphics[width=0.9\linewidth]{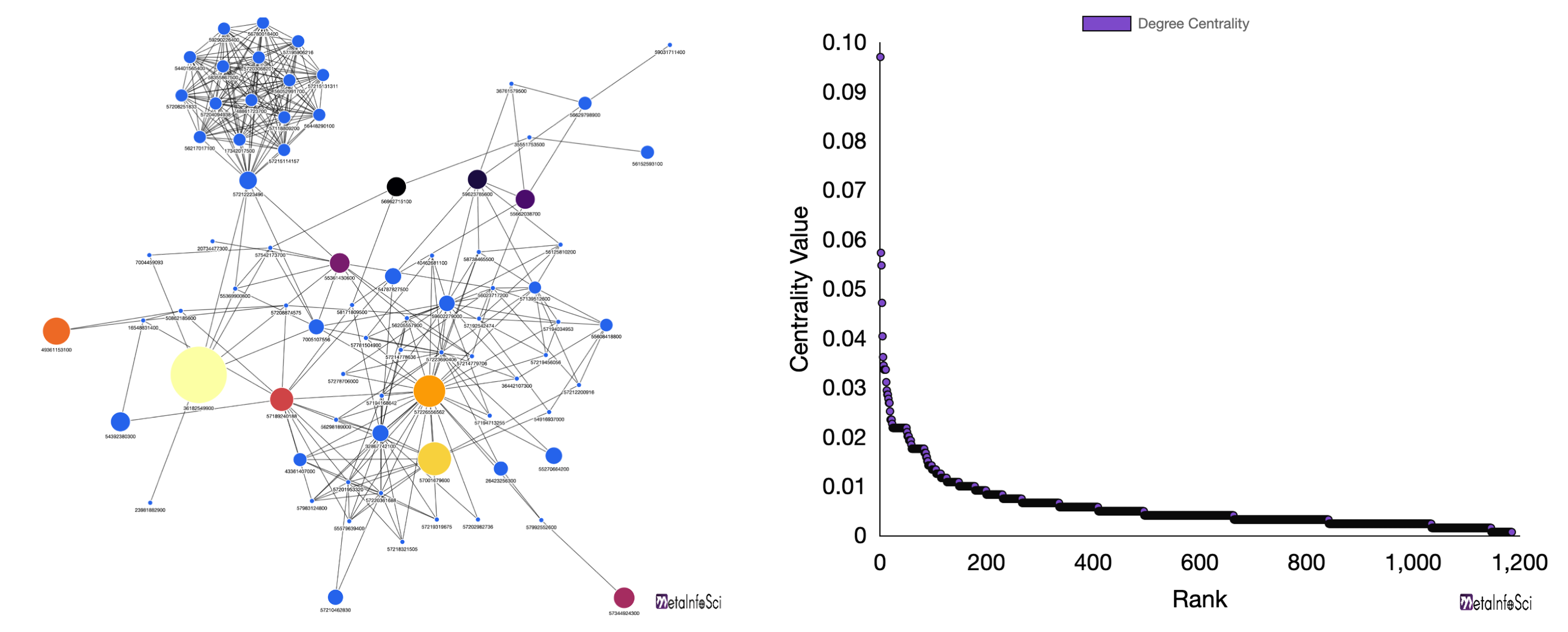} 
\caption{Degree centrality of author's collaboration network: (\textit{left}) Giant connected component of authors collaboration network with degree centrality. (\textit{right}) Probability distribution of degree centrality of each authors in giant connected component.}
\label{fig:R1-network}
\end{figure}

\begin{figure}[htbp]
    \centering
\includegraphics[width=0.9\linewidth]{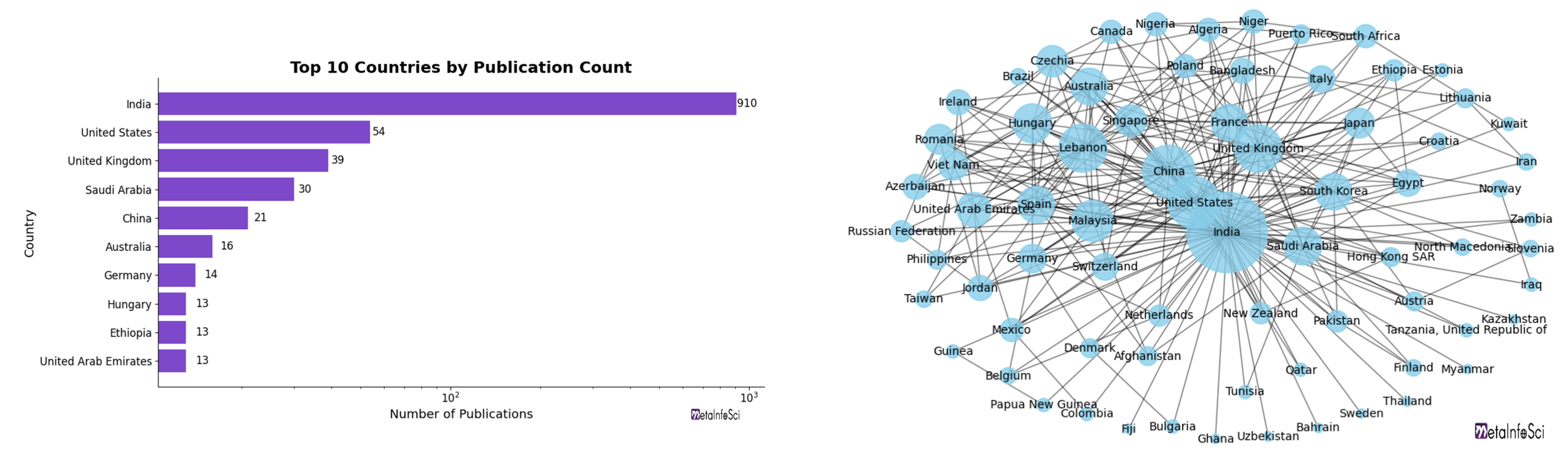} 
\caption{Country analysis: (\textit{left})  Top 10 countries as per number of publications. (\textit{right}) Country collaboration network with degree centrality.}
\label{fig:R1-Country}
\end{figure}
\begin{figure}[htbp]
    \centering
\includegraphics[width=0.68\linewidth]{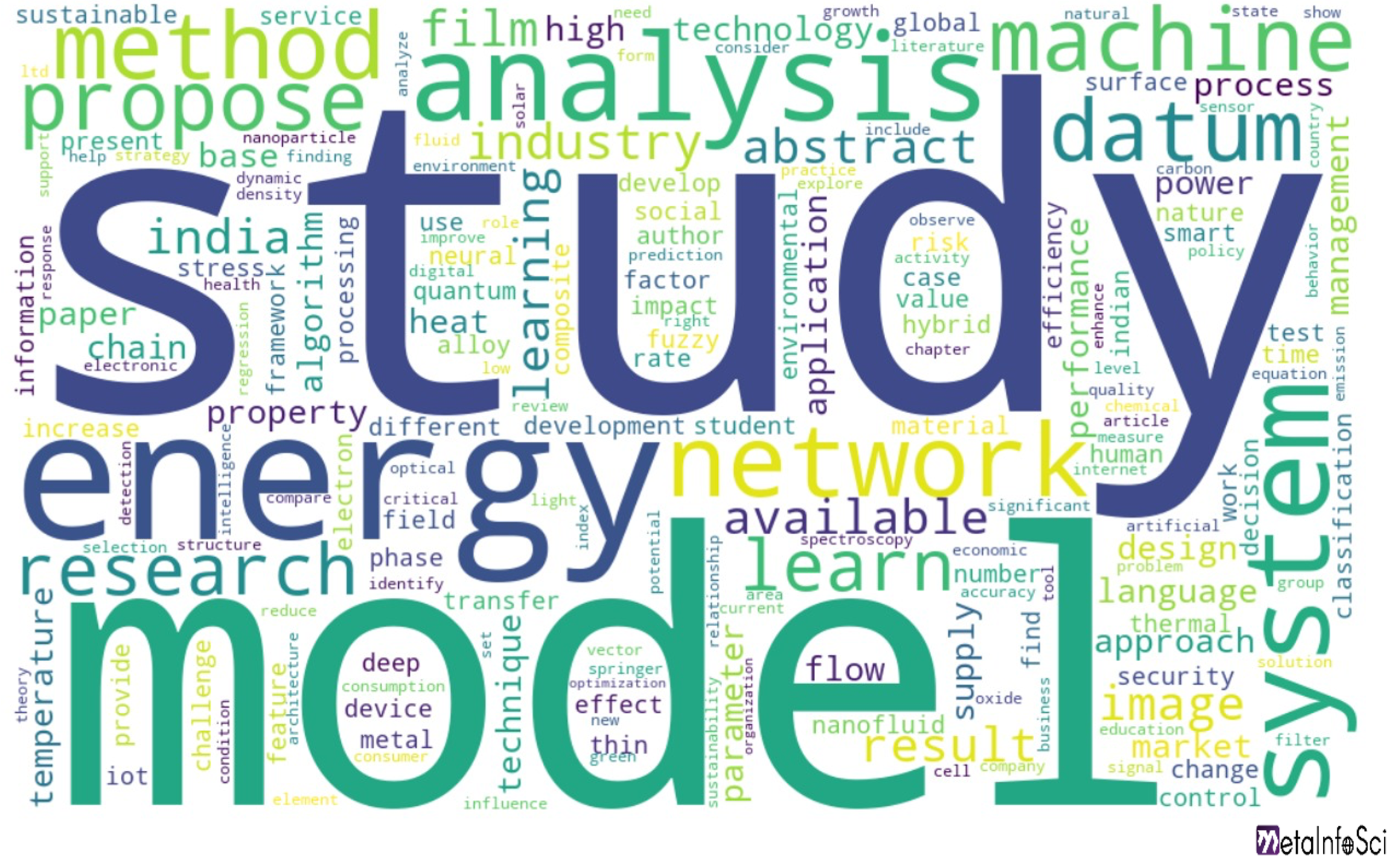} 
\caption{Keywords analysis: Word cloud of keywords.}
\label{fig:R1-WC}
\end{figure}


\section{Who Can Use MetaInfoSci and How They Benefit?}
\begin{enumerate}
    \item \textbf{Researchers and Academics}: MetaInfoSci empowers individual researchers across disciplines to explore scholarly trends, identify key literature, assess their impact, and visualize collaborations. Whether preparing a grant proposal, journal article, or thesis, researchers can make evidence-based decisions and communicate their work more effectively using intuitive data visualizations and AI summaries.
    
    \item \textbf{Universities and Research Institutions}: Academic institutions can use MetaInfoSci to benchmark their research performance, monitor departmental productivity, and visualize internal and external collaborations. This supports strategic planning, accreditation processes, and visibility in national and international rankings, making it a valuable tool for administrators and research offices.
    
    \item \textbf{Policymakers and Government Agencies}: Governments and science policy bodies can use MetaInfoSci to analyze national research output, assess contributions to societal challenges, and plan funding distribution. The platform aids evidence-based policymaking by identifying high-impact research areas and regional disparities in research activity, aligning scientific development with public needs.


    \item \textbf{Librarians and Research Support Staff}: University librarians and research support professionals can use MetaInfoSci to assist faculty and students in managing scholarly data, selecting suitable journals, and preparing for research assessments. The platform simplifies bibliometric services, making it easier to support academic publishing and compliance reporting.

    \item \textbf{Students and Early-Career Researchers}: MetaInfoSci serves as a learning and discovery tool for undergraduate and graduate students, helping them navigate academic literature and build well-informed theses or research proposals. It demystifies complex bibliometric data and supports the development of research literacy among the next generation of scholars.
\end{enumerate}

%
\section*{Acknowledgment}
Authors gratefully acknowledges the Research and Development Cell, BML Munjal University for their financial support through the seed grant (No: BMU/RDC/SG/2024-06), which made this research possible. We also appreciate Kothapalli Sravyanth, Aakarsh Goyal, Sairam Aditya Kasarla, Daksh Goel, and Kanhaiya Jha whose time to time contribution made this work possible.

 \section*{Conflict of interest}
 The author declares no conflict of interest.

\bibliographystyle{cas-model2-names}

\bibliography{cas-refs}
\end{document}